\documentclass[12pt]{article}

\usepackage{epsfig}

\usepackage{amsmath}
\usepackage{amssymb}
\usepackage{euscript}

\usepackage{cite}

\usepackage{alltt}

\newcommand{\Pcal}{{\cal P}}
\newcommand{\Peu}{\EuScript{P}}

\newcommand{\Pbf}{\mathbf{P}}

\newcommand{\from}{\leftarrow}

\def\half{{\textstyle{\frac{1}{2}}}}
\def\qi{{q_{i}}}
\def\qj{{q_{j}}}
\def\qibar{{\overline{q}_{i}}}
\def\qjbar{{\overline{q}_{j}}}
\def\qbar{{\overline{q}}}

\def\half{\frac{1}{2}}
\def\alfapi{\frac{\alpha_s}{\pi}}
\def\alfapi_t{\frac{\alpha_s}{2\pi}}
\def\Pbold{{\mathbf{P}}}
\def\Abold{{\mathbf{A}}}
\def\Bbold{{\mathbf{B}}}
\def\Cbold{{\overline{\mathbf{P}}}}
\def\Dbold{{\mathbf{D}}}
\def\Qbold{{\mathbf{Q}}}

\begin{document}

\begin{titlepage}


\begin{flushright}
\bf IFJPAN-V-04-08
\end{flushright}

\vspace{1mm}
\begin{center}
  {\LARGE\bf%
    Markovian Monte Carlo solutions \vspace{4mm}\\
    of the NLO QCD evolution equations$^{\star}$
}
\end{center}
\vspace{5mm}

\begin{center}
{\large\bf K.\ Golec-Biernat$^{a,c}$, S.\ Jadach$^a$, W.\ P\l{}aczek$^b$}
{\rm and}
{\large\bf M.\ Skrzypek$^a$} \\

\vspace{4mm}
{\em $^a$Institute of Nuclear Physics, Polish Academy of Sciences,\\
  ul.\ Radzikowskiego 152, 31-342 Cracow, Poland.}\\ \vspace{2mm}
{\em $^b$Marian Smoluchowski Institute of Physics, Jagiellonian University,\\
   ul.\ Reymonta 4, 30-059 Cracow, Poland.}\\ \vspace{2mm}
{\em $^c$Institute of Physics, University of Rzeszow,\\
   ul.\ Rejtana 16A, 35-959 Rzeszow, Poland.}
\end{center}

\vspace{15mm}
\begin{abstract}
We present precision Monte Carlo calculations solving the QCD evolution 
equations up to the next-to-leading-order (NLO) level. 
They employ forward Markovian Monte Carlo (FMC) algorithms,
which provide the rigorous solutions of the QCD evolution equations.
Appropriate Monte Carlo algorithms are described in detail.
They are implemented in the form of the Monte Carlo program {\tt EvolFMC}, 
which features the NLO kernels for the QCD evolution.
The presented numerical results agree with those from independent, 
non-MC, programs ({\tt QCDNum16}, {\tt APCheb33}) at the level of $0.1\%$.
In this way we have demonstrated the feasibility of the precision MC 
calculations for the QCD evolution 
and provided very useful numerical tests (benchmarks)
for other, non-Markovian, MC algorithms developed recently.
\end{abstract}

\vspace{15mm}
\begin{center}
\em To be submitted to Acta Physica Polonica B
\end{center}

\vspace{15mm}
\begin{flushleft}
{\bf IFJPAN-V-04-08}
\end{flushleft}

\vspace{5mm}
\footnoterule
\noindent
{\footnotesize
$^{\star}$This work is partly supported by the EU grant MTKD-CT-2004-510126
 in partnership with the CERN Physics Department and by the Polish Ministry
 of Scientific Research and Information Technology grant No 620/E-77/6.PR
 UE/DIE 188/2005-2008.
}

\end{titlepage}

\section{Introduction}
\label{sec:intro}

It is commonly known that the so-called evolution equations of 
the quark and gluon distributions in the hadron,
derived in QED and QCD using the renormalization group or diagrammatic
techniques~\cite{DGLAP},
can be interpreted probabilistically as a Markovian process,
see e.g.\ Ref.~\cite{stirling-book}.
Such a process can be modeled using Monte Carlo (MC) methods.
The corresponding MC algorithm, called in the following the Markovian
MC, provides, in principle, an exact solution of the evolution
equations for parton distribution functions (PDFs). In practice, the
main limitation of a such solution is the size
of a generated MC sample, i.e. corresponding statistical errors of numerical
results. This is probably the main reason why this possibility
has not been exploited until recently. 
Instead, alternative numerical methods and
programs solving the QCD evolution equations much faster than the Markovian
MC have been used. Typical examples of such non-MC programs are 
{\tt QCDNum16}~\cite{qcdnum16} and {\tt APCheb33}~\cite{APCheb33}, see also 
Ref.~\cite{blumleinn96}. 

Feasibility of solving efficiently the DGLAP equations~\cite{DGLAP} 
at the leading-order (LO) approximation with the Markovian MC was demonstrated 
for the first time by two of us (S.J. and M.S.) in Ref.~\cite{Jadach:2003bu}. 
There, the basic formalism was briefly sketched and the first numerical results
were presented. Good agreement between the constructed Markovian MC program
and {\tt QCDNum16} for gluon and quark-singlet distribution functions
was achieved. However, some small residual differences,  
at the level of $0.2\%$, between the two programs were found. 
Their origin was not understood at that time.
Here we repeat the above comparisons, explain the source of these discrepancies
and show the corrected results which agree at the level of $0.1\%$.
The main conclusion of Ref.~\cite{Jadach:2003bu} was that the currently
available computer CPU power allows to solve efficiently and precisely
(at the per-mill level) the QCD evolution equations with the use of the
Markovian MC algorithm. Of course, this method will always be slower in 
CPU time than non-MC techniques. However, it has several advantages, such as: 
no biases and/or numerical instabilities related to finite grids of points, 
use of quadratures, decomposition into finite series of polynomials,
accumulation of rounding errors, etc. It is also more flexible in 
treatment of PDFs (e.g. no need to split them into singlet and non-singlet 
components) and easier to extend into higher orders, new contributions, etc.

The above Markovian algorithm can be a basis for the final-state
radiation (FSR) parton shower MC program that not only solves numerically
the evolution equations but also generates events in terms of parton
flavours and four-momenta. Moreover, this algorithm can be a starting 
point and a testing tool for various kinds of constrained MC 
algorithms~\cite{raport04-06,raport04-07,zinnowitz04,herwig-cmc}
being developed for the initial-state radiation (ISR).

This paper is devoted to the Markovian MC solution of the DGLAP 
evolution equations up to the next-to-leading order (NLO) 
in the perturbative QCD.
It is organized as follows: In Section~2 we present a general structure
of the DGLAP equations and discuss their basic features up to the
next-to-next-to-leading order (NNLO). 
In Section~3 we describe in detail the Markovian MC algorithm for
parton density distributions. We start from a classic iterative solution
of the DGLAP equations and show how it can be expressed in terms of
Markovian transition probabilities. Then we provide a method for generation 
of a single Markovian step according to these  probabilities.  
Subsection~3.4 is devoted to construction of a weighted Markovian MC 
algorithm, where some importance sampling is used to generate evolution 
variables. In the last subsection we show how the above algorithm 
can be modified in order to account for the running QCD coupling constant.
In Section~4 we present the Markovian MC algorithm for parton-momentum
distributions. It has certain advantages over the previous algorithm
due to momentum sum rules that can be applied to evolution kernels.
Both the above algorithms have been implemented in the MC program 
called {\tt EvolFMC}~\cite{EvolFMC:2005}.
Numerical results from {\tt EvolFMC} at the LO and the NLO are presented 
in Section~5. They are compared with the results of non-MC programs
{\tt QCDNum16} and {\tt APCheb33}.
Section~6 summarizes the paper and gives some outlook for the future. 
In Appendix~A we collect formulae for the QCD kernels (splitting functions)
up to the NLO as well as explicit expressions for the NLO Sudakov 
form-factor. Appendices~B and C
contain formulae for simplified evolution kernels that are used for
importance sampling in the weighted Markovian algorithm.
Finally, in Appendix~D we discuss a generic discrete Markovian process. 
It can be seen as an illustration of basic features 
of the DGLAP-like evolution equations and their solution in terms of the
Markovian MC algorithm.

\section{QCD evolution equations}
\label{sec:QCD-evolution}

\subsection{General structure of DGLAP  equations} 
\label{sec:dglap}

The DGLAP evolution equations for  quark,
antiquark and gluon distributions,
\begin{equation}\label{eq:dglap1}
\{q_1,\ldots , q_{n_f},~ \qbar_1,\ldots ,\qbar_{n_f},~ G\}(\mu,x)\,,
\end{equation} 
take the following general form \cite{DGLAP,Furmanski:1981cw}
\begin{eqnarray}\nonumber
\label{eq:dglap2}
\frac{\partial}{\partial\ln\mu^2}\,\qi\!\!&=&\!\!\sum_j\left(
P_{\qi\qj}\otimes \qj\, +\, P_{\qi\qjbar}\otimes \qjbar\right) +\, P_{\qi G}\otimes G
\\\nonumber
\\\nonumber
\frac{\partial}{\partial\ln\mu^2}\,\qibar\!\!&=&\!\!\sum_j\left(
P_{\qibar\qj}\otimes \qj\, +\, P_{\qibar\qjbar}\otimes \qjbar\right) +\, P_{\qibar G}\otimes G
\\\nonumber
\\
\frac{\partial}{\partial\ln\mu^2}\, G\!\!&=&\!\!\sum_j\left(
P_{G\qj}\otimes \qj\, +\, P_{G\qjbar}\otimes \qjbar\right)
+\, P_{G G}\otimes G
\end{eqnarray}
where the summation is performed over quark flavours, $j=1,\ldots, n_f$.
The parton distributions are functions of the Bjorken variable $x$ and the
factorization scale $\mu$, identified with a hard scale
in a given process (e.g. ${\mu=\sqrt{Q^2}}$ in deep inelastic scattering). 
The functions $P=P(\mu,x)$ are splitting functions to be discussed below.
The integral convolution denoted by $\otimes$ involves only 
{\it longitudinal} momentum fractions
\begin{eqnarray}\nonumber
\label{eq:dglap3}
(P\otimes q)(\mu,x)\!\!\!&=&\!\!\!
\int\limits_0^1dy \int\limits_0^1dz\,\delta(x-zy)\,P(\alpha_s,z)\, q(\mu,y)
\\\nonumber
\\
&=&\!\!\!
\int\limits_x^1\frac{dz}{z} \,P(\alpha_s,z)\,q\left(\mu,\frac{x}{z}\right)
\,=\,\int\limits_x^1\frac{dz}{z}\,P\left(\alpha_s,\frac{x}{z}\right)\,q(\mu,z)
\,.
\end{eqnarray}
The splitting functions $P(\alpha_s,z)$ depend on $\mu$ through the strong coupling 
constant $\alpha_s=\alpha_s(\mu)$:
\begin{eqnarray}\label{eq:dglap6}
P(\alpha_s,z)\,=\,\frac{\alpha_s}{2\pi}\,P^{(0)}(z)\,+\,
\left(\frac{\alpha_s}{2\pi}\right)^2 P^{(1)}(z)\,+\,
\left(\frac{\alpha_s}{2\pi}\right)^3 P^{(2)}(z)\,+\ldots\,.
\end{eqnarray}
The superscripts $(0),(1),(2)$  refer  respectively  to the leading (LO), next-to-leading (NLO) and next-to-next-to-leading order (NNLO) approximations in which the splitting functions are computed\footnote{We adopt the convention of Curci, Furmanski 
            and Petronzio~\cite{Curci:1980uw,Furmanski:1980cm} in which the expansion parameter equals $\alpha_s/(2\pi)$. 
            The  NNLO analysis of Moch and 
            Vogt~\cite{Moch:2004pa,Vogt:2004mw} 
            uses $\alpha_s/(4\pi)$.}.

From charge conjugation and $SU(n_f)$ symmetry the splitting functions $P$
have the following general structure which is independent of the approximation
in which they have been computed
\begin{eqnarray}\label{eq:dglap4}
\nonumber
P_{\qi\qj}\!\!&=&\!\!P_{\qibar\qjbar}\,=\,\delta_{ij}P_{qq}^V\,+\,P_{qq}^S
\\\nonumber
P_{\qi\qjbar}\!\!&=&\!\!P_{\qibar\qj}\,=\,\delta_{ij}P_{q\qbar}^V\,+\,P_{q\qbar}^S
\\\nonumber
P_{\qi G}\!\!&=&\!\!P_{\qibar G}\,=\,P_{FG}
\\
P_{G\qi}\!\!&=&\!\!P_{G\qibar}\,=\,P_{GF}\,.
\end{eqnarray}
Substituting these relations to (\ref{eq:dglap2}), we find
\begin{eqnarray}\nonumber
\label{eq:dglap5}
\frac{\partial}{\partial\ln\mu^2}\, \qi\!\!&=&\!\!
P_{qq}^V\otimes\,\qi\,+\,P_{q\qbar}^V\otimes\,\qibar
\,+\,P_{qq}^S\otimes\sum_j\qj\,+\,P_{q\qbar}^S\otimes\sum_j\qjbar
\, +\, P_{F G}\otimes G
\\\nonumber
\\\nonumber
\frac{\partial}{\partial\ln\mu^2}\, \qibar\!\!&=&\!\!
P_{q\qbar}^V\otimes\,\qi\,+\,P_{qq}^V\otimes\,\qibar\,+\,
P_{q\qbar}^S\otimes\sum_j\qj\,+\,P_{qq}^S\otimes\sum_j\qjbar
\, +\, P_{F G}\otimes G
\\\nonumber
\\
\frac{\partial}{\partial\ln\mu^2}\, G\!\!&=&\!\!P_{GF}\otimes \sum_j(\qj+\qjbar)\, +\, P_{G G}\otimes G
\end{eqnarray}
This is the basic form of the DGLAP evolution equations.

Within a given  approximation  some splitting functions may
vanish  or be equal. In particular,
\begin{itemize}
\item in the LO \cite{DGLAP}
\begin{equation}\label{eq:dglap7a}
P^{V(0)}_{q\qbar}\,=\,P^{S(0)}_{q\qbar}\,=\,P^{S(0)}_{qq}\,=\,0\,,
\end{equation}
\item  in the NLO \cite{Curci:1980uw,Furmanski:1980cm}
\begin{equation}\label{eq:dglap7b}
P_{qq}^{S(1)}\,=\,P_{q\qbar}^{S(1)}\,,
\end{equation}
\item  but in the NNLO \cite{Moch:2004pa,Vogt:2004mw}
\begin{equation}\label{eq:dglap7c}
P_{qq}^{S(2)}\,\ne\,P_{q\qbar}^{S(2)}.
\end{equation}
\end{itemize}

Eqs.~(\ref{eq:dglap5}) can be rewritten in an alternative form which involves quark singlet and non-singlet distributions. We will present this form below.

\subsubsection{Singlet case}
The quark singlet  distribution is defined as 
\begin{equation}\label{eq:dglap8}
\Sigma(\mu,x)\,=\,\sum_{j=1}^{n_f}\left(q_j(\mu,x)+\qbar_j(\mu,x) \right)\,.
\end{equation}
Performing summation over quark flavours in the first two equations 
(\ref{eq:dglap5}),  we find
\begin{equation}
\label{eq:dglap9}
\frac{\partial}{\partial\ln\mu^2}\, \Sigma\,=\,
\left\{P_{qq}^V+P_{q\qbar}^V+n_f (P_{qq}^S+P_{q\qbar}^S)\right\}\otimes\, \Sigma
\, +\,(2n_f P_{FG})\otimes G
\end{equation}
Introducing the notation
\begin{eqnarray}
\label{eq:dglap10a}
P_{FF}\!\!\!&=&\!\!\!P_{+}^{V}+n_f P_{+}^{S}
\\\nonumber
\\
\label{eq:dglap10b}
P^{V,S}_{\,+}\!\!\!&=&\!\!\!P^{V,S}_{qq}\,+\, P^{V,S}_{q\qbar}\,,
\end{eqnarray}
the following closed set of equations is obtained for the quark singlet 
and gluon distributions
\begin{eqnarray}
\label{eq:dglap11a}
\frac{\partial}{\partial\ln\mu^2}\, \Sigma\!\!&=&\!\!
P_{FF}\otimes\, \Sigma\, +\,(2n_f P_{F G})\otimes G
\\\nonumber
\\
\label{eq:dglap11b}
\frac{\partial}{\partial\ln\mu^2}\, G\!\!&=&\!\!P_{GF}\otimes \Sigma\, +\, P_{GG}\otimes G\,.
\end{eqnarray}

 The splitting functions in these equations obey the general relations
\begin{equation}\label{eq:dglap12}
\int\limits_0^1 dz\,\{\,zP_{FF}(\mu,z)+zP_{GF}(\mu,z)\,\}\,=\,
\int\limits_0^1 dz\,\{\,2n_f zP_{F G}(\mu,z)+zP_{GG}(\mu,z)\,\}\,=\,0\,.
\end{equation}
They immediately leads to  the momentum sum rule  
\begin{equation}\label{eq:dglap13}
\int\limits_0^1dx \left\{x\Sigma(\mu,x)\,+\,xG(\mu,x)\right\}\,=\,{\rm const}\,.
\end{equation}
which is conserved during the evolution. In the parton model interpretation 
the constant is set to one by normalizing the initial
conditions for Eqs.~(\ref{eq:dglap11a},\ref{eq:dglap11b}): 
$\Sigma(\mu_0,x)$ and $G(\mu_0,x)$.

\subsubsection{Non-singlet case}
Introducing the quark non-singlet distribution
\begin{equation}\label{eq:dglap14}
V(\mu,x)\,=\,\sum_{j=1}^{n_f}\, (q_j(\mu,x)-\qbar_j(\mu,x))\,,
\end{equation}
the following evolution equation is obtained from Eqs.~(\ref{eq:dglap5})
\begin{equation}\label{eq:dglap15}
\frac{\partial}{\partial\ln\mu^2}\,V\,=\,P^V_{NS}\otimes V\,,
\end{equation}
where the new splitting function reads
\begin{eqnarray}
\label{eq:dglap16a}
P^V_{NS} \!\!\!&=&\!\!\! P^{V}_{\,-}+n_fP^S_{\,-}
\\\nonumber
\\
\label{eq:dglap16b}
P^{V,S}_{\,-}\!\!\!&=&\!\!\!  P^{V,S}_{qq}-P^{V,S}_{q\qbar}\,.
\end{eqnarray}
Similarly, for  the non-singlet quark distributions
\begin{eqnarray}\label{eq:dglap17}
q_i^{-}(\mu,x)\!\!\!&=&\!\!\!
q_i(\mu,x)-\qbar_i(\mu,x)-\frac{1}{n_f}V(\mu,x)
\\\nonumber
\\
q_i^{+}(\mu,x)\!\!\!&=&\!\!\!
q_i(\mu,x)+\qbar_i(\mu,x)-\frac{1}{n_f}\Sigma(\mu,x)\,,
\end{eqnarray}
we find from Eqs.~(\ref{eq:dglap5}), (\ref{eq:dglap11a})  and (\ref{eq:dglap15}) the following
equations
\begin{eqnarray}\label{eq:dglap18a}
\frac{\partial}{\partial\ln\mu^2}q_i^{-}\!\!\!&=&\!\!\!P^{V}_{\,-}\otimes q_i^{-}
\\\nonumber
\\\label{eq:dglap18b}
\frac{\partial}{\partial\ln\mu^2}q_i^{+}\!\!\!&=&\!\!\!P^{V}_{\,+}\otimes q_i^{+}\,.
\end{eqnarray}
Notice that there is no gluon distribution in the derived equations.

\subsection{Summary of the forms}
\label{sec:sumary}
With the splitting functions usually presented in the literature
\begin{equation}\label{eq:dglap22}
\{P^{V}_{\pm},\,P^S_{\pm},\,P_{FG},\,P_{GF},\,P_{GG}\}\,,
\end{equation}
the evolution equations for the parton distributions  $\{q_i^{-},\,q_i^{+},V,\,\Sigma,\,G\}$
are given  by  Eqs.~(\ref{eq:dglap18a}), (\ref{eq:dglap18b}), 
(\ref{eq:dglap15}), (\ref{eq:dglap11a}) and (\ref{eq:dglap11b}), respectively.

According to relations (\ref{eq:dglap7a})--(\ref{eq:dglap7c}), the perturbative
expansions for the kernels $P^{V}_{\pm~}$ and $P^{S}_{\pm~} $ take the following form
\begin{eqnarray}\nonumber
\label{eq:dglap24a}
P^{V}_{\pm~}(\alpha_s,z)\!\!\!&=&\!\!\!
\alfapi_t\, P^{V(0)}_{qq}(z)\,
+\left(\alfapi_t\right)^2 P^{V(1)}_{\pm}(z)\,
+\left(\alfapi_t\right)^3 P^{V(2)}_{\pm}(z)\,+\,\ldots
\\\nonumber
\\\nonumber
P^{S}_{+~}(\alpha_s,z)\!\!\!&=&\!\!\!
~~~~~~~~~~~~~~~~~~~
\left(\alfapi_t\right)^2 P^{S(1)}_{qq}(z)\,\,
+\left(\alfapi_t\right)^3 P^{S(2)}_{+~}(z)\, +\,\ldots
\\\nonumber
\\
P^{S}_{\,-~}(\alpha_s,z)\!\!\!&=&\!\!\!
~~~~~~~~~~~~~~~~~~~~~~~~~~~~~~~~~~~~~~~~~~~~
\left(\alfapi_t\right)^3 P^{S(2)}_{\,-~}(z)\, +\,\ldots\,
\end{eqnarray}
The remaining kernels  $\{P_{FG},\,P_{GF},\,P_{GG}\}$  have the  nonzero splitting functions
in each approxi\-ma\-tion. 

Alternatively, the parton distributions
$\{q_i,\,\qbar_i,\,G\}$ could be evolved with the help of  Eqs.~(\ref{eq:dglap5}) 
with the kernels
\begin{equation}\label{eq:dglap25}
\{P_{qq}^{V,S},\,P_{q\qbar}^{V,S},\,P_{FG},\,P_{GF},\,P_{GG}\}\,,
\end{equation}
where $P^{V,S}_{qq}$ and $P^{V,S}_{q\qbar}$ are computed 
by inverting relations (\ref{eq:dglap10b}) and (\ref{eq:dglap16b}):
\begin{eqnarray}
\label{eq:dglap26a}
P_{qq}^{V,S}\!\!\!&=&\!\!\! 
\textstyle{\half} \left(P^{V,S}_{+}+P^{V,S}_{\,-}\right)
\\\nonumber
\\
\label{eq:dglap26b}
P_{q\qbar}^{V,S} \!\!\!&=&\!\!\!
\textstyle{\half} \left(P^{V,S}_{+}-P^{V,S}_{-}\right).
\end{eqnarray}

\subsection{Behaviour at $z\to 1$}
\label{sec:z1}
Let us consider the splitting functions (\ref{eq:dglap22}).  All the kernels, except the $P^S_{+}$,
are divergent for $z=1$.

The quark-quark and gluon-gluon splitting functions
$
\{P^V_{\pm},\,P^S_{\,-},\,P_{GG}\}
$
have  the following form
\begin{equation}\label{eq:dglap27}
P(\alpha_s,z)\,=\,\frac{A(\alpha_s)}{(1-z)_+}\,+\,
B(\alpha_s)\,\delta(1-z)\,+\,\overline{P}(\alpha_s,z)\,,
\end{equation}
where the ``$+$'' prescription regularizes the $1/(1-z)$ singularity
\begin{equation}\label{eq:dglap27a}
[f(z)]_+\,=\,f(z)\,-\,\delta(1-z)\int\limits_0^1dz^\prime f(z^\prime)\,.,
\end{equation}
and 
the functions $A(\alpha_s),\,B(\alpha_s)$ and $\overline{P}(\alpha_s,z)$ are computed
in powers of $\alpha_s$, see Eq.~(\ref{eq:dglap6}). 
In particular
\begin{equation}
\overline{P}(\alpha_s,z)\,=\,\sum_{k=0}\alpha_s^{k+1}\,{D}^{(k)}(z)\,.
\end{equation}
In the LO approximation $(k=0)$ ${D}^{(0)}(z=1)$ is finite \cite{DGLAP} while in the
NLO $(k=1)$ and NNLO $(k=2)$ approximations,  the coefficients
are logarithmically divergent
\cite{Curci:1980uw,Furmanski:1980cm,Moch:2004pa,Vogt:2004mw}:
\begin{equation}
{D}^{(k)}(z)\,=\,D_k\,\ln(1-z)\,+\,{\cal O}(1)\,.
\end{equation}

Similarly, 
the quark-gluon and gluon-quark splitting functions
$
\{P_{FG},\,P_{GF}\}
$
contain logarithmically divergent terms for $z=1$ \cite{Furmanski:1980cm,Vogt:2004mw}:
\begin{equation}
\label{eq:dglap28}
P(\alpha_s,z)\,=\,\sum_{k=0}\alpha_s^{k+1}\bigg\{
\sum_{i=1}^{2k}\, D_i^{(k)}\,\ln^i(1-z)\,+\,{\cal O}(1)\bigg\}\,,
\end{equation}
Thus in the limit $z\to 1$, we have for $P_{FG}$ and $P_{GF}$:
\\
\begin{equation}
P(\alpha_s,z)\,=\,
\left\{
\begin{array}{ll}
{\cal{O}}(\alpha_s)~~~~~                          &     {\rm in~LO~(k=0)}\\  \\
{\cal{O}}(\alpha_s^2\ln^2(1-z))~~~~~              &     {\rm in~NLO~(k=1)}\\  \\
{\cal{O}}(\alpha_s^3\ln^4(1-z))~~~~~              &     {\rm in~NNLO~(k=2).}
\end{array}
\right.
\end{equation}

\subsection{Behaviour at $z\to 0$}
As in the previous section, let us
consider the splitting functions (\ref{eq:dglap22}).

The splitting functions
$
\{P^V_{\pm},P^S_{\,-}\}
$
are logarithmically divergent at $z=0$ starting from the NLO approximation: \cite{Curci:1980uw,Furmanski:1980cm,Moch:2004pa,Vogt:2004mw}:
\begin{equation}\label{eq:dglap29a}
P(\alpha_s,z)\,=\,\sum_{k=0}\alpha_s^{k+1}\,
\bigg\{\sum_{i=1}^{2k}\,{\overline D}_i^{(k)}\,\ln^{i}z\,+\,{\cal O}(1)\bigg\},
\end{equation}
Thus for $z\to 0$, we find for $P^V_{\pm}$ and $P^S_{\,-}$:
\begin{equation}
P(\alpha_s,z)\,=\,
\left\{
\begin{array}{ll}
{\cal{O}}(\alpha_s)~~~~~                          &     {\rm in~LO~(k=0)}\\  \\
{\cal{O}}(\alpha_s^2\ln^2 z)~~~~~              &     {\rm in~NLO~(k=1)}\\  \\
{\cal{O}}(\alpha_s^3\ln^4 z)~~~~~              &     {\rm in~NNLO~(k=2).}
\end{array}
\right.
\end{equation}

The remaining splitting functions
$
\{P^S_{\,+},P_{FG},\,P_{GF},\,P_{GG}\}
$
have the following behaviour  for $z\to 0$ \cite{Furmanski:1980cm,Vogt:2004mw}:
\begin{equation}\label{eq:dglap29b}
P(\alpha_s,z)\,=\,E_1(\alpha_s)\,\frac{\ln z}{z}\,+\,E_2(\alpha_s)\,\frac{1}{z}\,+\,
{\cal{O}}(\ln^{2k}\!z )\,,
\end{equation}
The logarithmic term is present starting from the NLO $(k=1)$ approximation:
\begin{equation}
E_1(\alpha_s)\,=\,\alpha_s^2\,E_1^{(1)}+\,\alpha_s^3\,E_1^{(2)}\,+\,...\,,
\end{equation}
while the $1/z$ term is present from the LO $(k=0)$ approximation:
\begin{equation}
E_2(\alpha_s)\,=\,\alpha_s\,E_2^{(0)}+\,\alpha_s^2\,E_2^{(1)}+\,\alpha_s^3\,E_2^{(2)}...\,.
\end{equation}

\subsection{Monte Carlo form of DGLAP equations}
The $z=1$ singularity in Eq.~(\ref{eq:dglap27}) needs special treatment 
in the Monte Carlo formulation of the DGLAP equations. Rewritting Eqs.~(\ref{eq:dglap5})
for the parton distributions multiplied by $x$, denoted in the matrix form as
\begin{equation}\nonumber
\label{eq:dglap30}
\left\{x\/q_1,\ldots , x\/q_{n_f},~ x\/\qbar_1,\ldots ,x\/\qbar_{n_f},~ x\/G  \right\}(\mu,x)\equiv \,x\Dbold(\mu,x)\equiv \Qbold(\mu,x)\,,
\end{equation}
we have
\begin{equation}
\label{eq:dglap31}
\frac{\partial}{\partial\ln\mu^2}\, \Qbold(\mu,x)\,=\,\int\limits_x^1
{dz}\,\Pbold(\alpha_s,z)\,\Qbold\left(\mu,\frac{x}{z}\right)\,,
\end{equation}
where $\Pbold$  is the matrix of the splitting functions, which can be easily read off from Eqs.~(\ref{eq:dglap5}).

Based on the results of  Section \ref{sec:z1}, we can write the general structure of the splitting
functions  in the following form
\begin{equation}
\label{eq:dglap32}
\Pbold(\alpha_s,z)\,=\,\frac{\Abold(\alpha_s)}{(1-z)_+}\,+\,
\Bbold(\alpha_s)\,\delta(1-z)\,+\,\Cbold(\alpha_s,z)
\end{equation}
where $A,\,B,\, \Cbold$ are computed in powers of $\alpha_s$. The function 
$\Cbold(\alpha_s,z)$ may contain singular terms in the limit $z\to 1$,
proportional to powers of $\ln(1-z)$.

For simplicity of the notation  we suppress the $\mu$-dependence of the parton distribution 
and the splitting functions in the following.
Substituting
(\ref{eq:dglap32}) to Eq.~(\ref{eq:dglap31}) and using definition (\ref{eq:dglap27a}), we find
\begin{eqnarray}\nonumber
\label{eq:dglap33}
\frac{\partial}{\partial\ln\mu^2}\, \Qbold(x)\!\!\!&=&\!\!\!\int\limits_x^1{dz}
\left\{
\Abold\,\frac{\Qbold(x/z)-\Qbold(x)}{1-z}\,+\,\Cbold(z)\,\Qbold(x/z)
\right\}
\\\nonumber
\\
&&~~~~
+\left\{\Abold\,\ln(1-x)+\Bbold\,\right\}\Qbold(x)\,.
\end{eqnarray}
Now, we introduce a small cutoff in the upper limit of the integration,
$1\to (1-\epsilon)$, which isolates the $z=1$ singularity. Performing the integration,
\begin{equation}\nonumber
\label{eq:dglap34}
\int\limits_x^{1-\epsilon}{dz}\,
\Abold\,\frac{-\Qbold(x)}{1-z}\,=\,
\left\{\Abold\ln\epsilon\,-\,\Abold\ln(1-x)\right\}\Qbold(x)\,,
\end{equation}
we find for Eq.~(\ref{eq:dglap33})
\begin{eqnarray}\nonumber
\label{eq:dglap35}
\frac{\partial}{\partial\ln\mu^2}\, \Qbold(x)=
\int\limits_x^{1-\epsilon}\!\!{dz}\,
\Abold\,\frac{\Qbold(x/z)}{1-z}
\,+\,
\int\limits_x^{1}{dz} \,\Cbold(z)\,\Qbold(x/z)
\\\nonumber
\\
\,\,+\,\left\{\Abold\ln\epsilon+\Bbold\right\}\Qbold(x)\,.
\end{eqnarray}

Inserting back the $\mu$-dependence, the equation above  can be written as
\begin{equation}
\label{eq:dglap36}
\frac{\partial}{\partial\ln\mu^2}\,\Qbold(\mu,x)\,=\int\limits_x^1dz\,\Pbold(\alpha_s,z,\epsilon)\,\Qbold(\mu,x/z)
\end{equation}
with the kernel
\begin{equation}
\label{eq:dglap37}
\Pbold(\alpha_s,z,\epsilon)\,=\,\frac{\Abold(\alpha_s)}{1-z}\,\Theta(1-z-\epsilon)
\,+\,\left\{\Abold(\alpha_s)\ln\epsilon+\Bbold(\alpha_s)\right\}\delta(1-z)
\,+\,\Cbold(\alpha_s,z)
\,.
\end{equation}

This form of the DGLAP equations is a starting point for the Monte Carlo
generation. Let us notice that the presented formulae
are valid for both representations of the parton distributions and splitting functions discussed in Section \ref{sec:sumary}.
Explicit expressions for the splitting functions up to the NLO are given 
in Appendix~\ref{sec:kernelsNLO}.

\section{Markovian algorithm for parton distributions}
\label{sec-multicomp}

In the following we show how to transform the QCD evolution equation (DGLAP 
type) into an integral homogeneous equation and solve it by means of iteration.
The general properties of the evolution equations and the related diffusion
equations are discussed in Appendix~\ref{sec:discrete}
using simple environment of the discrete space.
Below we discuss a more complicated case of the mixed, discrete-continuous,
space of the QCD evolution equations.

\subsection{Classic iterative solution}

Introducing the  variable 
\begin{equation}
t=\ln \mu\equiv \ln Q\,,
\end{equation}
the evolution equations
(\ref{eq:dglap31}) in the component form read
\begin{equation}
\begin{split}
  \label{eq:Evolu}
  \frac{\partial}{\partial t} D_K(t,x)
  &\,=\, \sum_J (\Peu_{KJ}\otimes D_J)(t,x)\\
  &= \sum_J \int\limits_x^1 \frac{d z}{z}\, \Peu_{KJ}(t,z) \, D_J\Big(t,\frac{x}{z} \Big)
\end{split}
\end{equation}
Notice that due to the definition of the evolution variable $t$, the splitting functions $\Peu_{KJ}$ 
are related to those from section \ref{sec:intro}, Eqs.~(\ref{eq:dglap6}), by
\begin{equation}
\Peu_{KJ}(t,z)\,=\,2\/P_{KJ}(\alpha_s(t),z)
\end{equation}
with a possible dependence of $\alpha_s(t)$ also on $z$ in order to accommodate
coherence effects.

The next important step is to introduce the infra-red (IR) cut $\epsilon$
and to isolate a part of the kernel $\Peu$
diagonal both in the parton (flavour) index and in the $z$-variable%
\footnote{All components proportional to $\delta(1-z)$ 
  reside in the diagonal part of the matrix $\Peu_{kj}$ anyway.}
\begin{equation}
  \begin{split}
  \Peu_{KJ}(t,z)&=-\Peu^{\delta}_{KK}(t,\epsilon(t))\, \delta_{KJ}\,\delta(1-z)
                 +\Peu^{\Theta}_{KJ}(t,z),
\\
\Peu^{\Theta}_{KJ}(t,z)&=\Peu_{KJ}(t,z)\,\Theta(1-z-\epsilon(t))\,\Theta(z-\epsilon'),
  \end{split}
\end{equation}
where we also introduced a facultative lower limit $\epsilon'$ on
$z$-variable, equivalent to the minimal global $x$ of the evolution.
Note that in the DGLAP case there is no reason for the IR regulator $\epsilon$
to be $t$-dependent. However, for applications in the context 
of parton-shower algorithms and of the CCFM equations~\cite{CCFM}
it is worthwhile to keep this option open. In any case we always assume that
$\epsilon$ and $\epsilon(t)$ are small.
A more detailed discussion of the LO kernels is given in 
Appendix~\ref{sec:kernels1}.
After the above splitting of the kernels 
the evolution equation becomes inhomogeneous
\begin{equation}
  \label{eq:Evol3}
  \frac{\partial}{\partial t} D_K(t,x)
  + \Peu^{\delta}_{KK}(t)\; D_K(t,x)
  =\sum_J (\Peu^{\Theta}_{KJ}\otimes D_J)(t,x).
\end{equation}
It is easily made again homogeneous
\begin{equation}
  \label{eq:evol3_homo}
  \begin{split}
    & e^{-\Phi_K(t,t_0)}  \frac{\partial}{\partial t} 
    \left(  e^{\Phi_K(t,t_0)} D_K(t,x) \right)
    = \sum_J (\Peu^\Theta_{KJ} \otimes D_J)(t,x),\\
    &\Phi_{K}(t,t_0) = \int\limits_{t_0}^t  dt'\;
       \Peu^\delta_{KK}(t',\epsilon(t'))\,,\\
  \end{split}
\end{equation}
and turned into an integral equation
\begin{equation}
 e^{\Phi_K(t,t_0)}   D_K(t,x)
 = D_K(t_0,x) 
  +\int\limits_{t_0}^t dt_1  e^{\Phi_K(t_1,t_0)}   
   \sum_j (\Peu^\Theta_{KJ}\otimes D_J)(t_1,x).
\end{equation}
Its another equivalent form, which is more convenient for iteration, reads
\begin{equation}
 D_K(t,x)
 = e^{-\Phi_K(t,t_0)} D_K(t_0,x) 
  +\int\limits_{t_0}^t dt_1  e^{-\Phi_K(t,t_1)}   
   \sum_J (\Peu^\Theta_{KJ}\otimes D_J)(t_1,x).
\end{equation}
The iteration of the above equation provides a solution in terms of
a series of integrals
\begin{equation}
  \label{eq:IterBasic}
  \begin{split}
  D_K&(t,x) = e^{-\Phi_K(t,t_0)} D_K(t_0,x)
  +\sum_{n=1}^\infty \;
   \sum_{K_0,\ldots,K_{n-1}}
      \prod_{i=1}^n \bigg[ \int\limits_{t_0}^t dt_i\; \Theta(t_i-t_{i-1}) \int\limits_0^1 dz_i\bigg]
\\&~~~~~~\times
      e^{-\Phi_K(t,t_n)}
      \int\limits_0^1 dx_0\;
      \prod_{i=1}^n 
          \bigg[ \Peu_{K_i K_{i-1}}^\Theta (t_i,z_i) 
                 e^{-\Phi_{K_{i-1}}(t_i,t_{i-1})} \bigg]
      D_{K_0}(t_0,x_0) \delta\big(x- x_0\prod_{i=1}^n z_i \big),
  \end{split}
\end{equation}
where $k_n\equiv k$.
At this point we have many options for the MC implementation
of the multidimensional integrals given by the above expression.
Quite generally, they can be divided into Markovian and non-Markovian
groups of the MC implementations.
In the following we shall describe solutions of the Markovian type.
However, it will be done such that the mechanism to switch
to a non-Markovian method will be as easy as possible.

\subsection{Markovianization}

Contrary to the evolution of the non-singlet PDF or 
of the singlet one-component PDF, 
in the most general case represented by the Eq.~(\ref{eq:IterBasic})
one cannot express its integrand
as an exact product of the Markovian single-step
probabilities, each normalized to $1$.
However, the general iterative solution of the evolution equation
in Eq.~(\ref{eq:IterBasic})
can be expressed in terms of the (unnormalized) transition density
\begin{equation}
  \Omega(t_i,x_i,K_i| t_{i-1},x_{i-1},K_{i-1})
  \equiv  \Theta(t_i-t_{i-1})\;
     \Peu_{K_iK_{i-1}}^\Theta (t_i,x_i/x_{i-1})\;
     e^{-\Phi_{K_{i-1}}(t_i,t_{i-1})}
\end{equation}
as follows
\begin{equation}
  \label{eq:EvolIter2}
  \begin{split}
  D_K&(t,x) = e^{-\Phi_K(t,t_0)} D_K(t_0,x)
  +\sum_{n=1}^\infty \;
   \sum_{K_0,\ldots,K_{n-1}}
      \int\limits_0^1 dx_0\;
      \prod_{i=1}^n \bigg[ \int\limits_{t_0}^t dt_i\; \int\limits_0^1 dz_i \bigg] \;
      e^{-\Phi_K(t,t_n)}
\\&~~~~~~\times
      \prod_{i=1}^n 
	  \Omega(t_i,x_i,K_i| t_{i-1},x_{i-1},K_{i-1})
      \delta\big(x- x_0\prod_{i=1}^n z_i \big)\;
      D_{K_0}(t_0,x_0).
  \end{split}
\end{equation}
The above expression looks almost as a product of Markovian transition probabilities,
except that $\Omega$ lacks a proper normalization
\begin{equation}
  \int\limits_{t_{i-1}}^\infty dt_i\; \int\limits_0^1 dz_i \sum_{K_i}
  \Omega(t_i,x_i,K_i| t_{i-1},x_{i-1},K_{i-1})=
  \int\limits_0^\infty d( T_{K_{i-1}}(t_{i},t_{i-1}))\; 
  e^{-\Phi_{K_{i-1}}(t_{i},t_{i-1})} \neq 1,
\end{equation}
where
\begin{equation}
    T_{K}(t,t_0)
    = \int\limits_{t_0}^t  dt'\; \int\limits_{0}^{1} dz \;
             \sum_J \Peu^\Theta_{JK}(t',z).
\end{equation}
The above problem cannot be cured by changing integration variables
or normalization of the PDFs.
On the other hand, one can define a properly normalized transition probability
\begin{equation}
  \begin{split}  
    &\omega(t_i,x_i,K_i| t_{i-1},x_{i-1},K_{i-1})
    \equiv  \Theta(t_i-t_{i-1})\;
    \Peu_{K_iK_{i-1}}^\Theta (t_i,x_i/x_{i-1})\;
    e^{-T_{K_{i-1}}(t_i,t_{i-1})},\\
    &\int\limits_{t_{i-1}}^\infty dt_i\; \int\limits_0^1 dz_i \sum_{K_i}
    \omega(t_i,x_i,K_i| t_{i-1},x_{i-1},K_{i-1})\equiv 1,
  \end{split}
\end{equation}
and express
\begin{equation}
  \Omega(t_i,x_i,K_i| t_{i-1},x_{i-1},K_{i-1})
  = e^{\Delta_{K_{i-1}}(t_i,t_{i-1})}
  \omega(t_i,x_i,K_i| t_{i-1},x_{i-1},K_{i-1})\,,
\end{equation}
where
\begin{equation}
  \Delta_{K}(t,t_0) = T_{K}(t,t_0)-\Phi_{K}(t,t_0)
   = \int\limits_{t_0}^t  dt'\; \int\limits_{0}^{1} dz \;
    \sum_J \Peu_{JK}(t',z),
\end{equation}
which is independent of the IR regulator $\epsilon(t)$.
This opens a way to the Monte Carlo algorithm
with weighted events, in which the Markovian algorithm
is based on the $\omega$ distributions 
and the correcting weight 
$w = \prod (\Omega/\omega)$
brings back the MC distributions to the original ones.

As usual, Markovianization cannot be accomplished without
adding one extra integration variable.
We do this starting from the identity
\begin{equation}
  \int\limits_{t}^\infty dt_i\; \int\limits_0^1 dz_i \sum_{K_i}
  \omega(t_i,x_i,K_i| t_{i-1},x_{i-1},K_{i-1})
   = e^{-T_{K_{i-1}}(t,t_{i-1})}\,,
\end{equation}
and then we add the $(n+1)$-th ``spill-over'' variables in the integrals using
\begin{equation}
  e^{\Delta_{K_{n}}(t,t_{n})}
  \int\limits_{t}^\infty dt_{n+1}\; \int\limits_0^1 dz_{n+1} \sum_{K_{n+1}}
  \omega( t_{n+1},x_{n+1},K_{n+1}| t_n,x_n,K_n)
   = e^{-\Phi_{K_{n}}(t,t_{n})}\,.
\end{equation}

Summarizing all the above discussion, we transform Eq.~(\ref{eq:IterBasic}) into
a new equivalent form
\begin{equation}
  \label{eq:MarkovianStandard}
  \begin{split}
  D_K&(t,x) = 
  e^{\Delta_{K}(t,t_{0})}
  \int\limits_{t_{1}>t} dt_{1} dz_{1} \sum_{K_{1}}
  \omega( t_{1},x_{1},K_{1}| t_0,x,K)\;\;
  D_K(t_0,x)
\\&
  +\sum_{n=1}^\infty \;\;
  \int\limits_0^1 dx_0\;
  \int\limits_{t_{n+1}>t} dt_{n+1} dz_{n+1} \sum_{K_{n+1}}\;\;
  \sum_{K_0,\ldots,K_{n-1}}\;\;
      \prod_{i=1}^n \int\limits_{t_i<t} dt_idz_i \;
\\&~~~~~~\times
   e^{\Delta_{K_{n}}(t,t_{n})}
   \omega( t_{n+1},x_{n+1},K_{n+1}| t_n,x_n,K_n)
\\&~~~~~~\times
   \prod_{i=1}^n 
      e^{\Delta_{K_{i-1}}(t_i,t_{i-1})}
      \omega(t_i,x_i,K_i| t_{i-1},x_{i-1},K_{i-1})
\\&~~~~~~\times
      \delta\big(x- x_0\prod_{i=1}^n z_i \big)\;
      D_{K_0}(t_0,x_0).
  \end{split}
\end{equation}
In the Markovian Monte Carlo algorithm implementing exactly 
the above series of the integrals
we neglect primarily the factor
\begin{equation}
  \label{eq:wt_delta}
  w=e^{\Delta_{K_{n}}(t,t_{n})} 
      \prod_{i=1}^n e^{\Delta_{K_{i-1}}(t_i,t_{i-1})}\,,
\end{equation}
such that
the whole series of integrals can be implemented readily as a
Markovian chain of steps with the normalized
transition probability $\omega(t_i,x_i,K_i| t_{i-1},x_{i-1},K_{i-1})$
for each single step.
The original integrals and distributions can be recovered by means 
of applying the MC correcting weight $w$ defined above.
The only technical problem is that $w \geq 1$ and 
one may struggle to find the maximum weight,
in order to turn weighted events into unweighted ones.
It is an unavoidable price to pay in this method.

In the case of the single-component evolution (singlet or non-singlet)
we recover automatically the constant-weight algorithm
\begin{equation}
  w=e^{\Delta(t,t_{n})} 
    \prod_{i=1}^n e^{\Delta(t_i,t_{i-1})}
   =e^{\Delta(t,t_{0})}.
\end{equation}
In the case of the non-singlet evolution we even have $w=1$.

\subsection{Generation of a single Markovian step}

The description of the Markovian algorithm of the previous section
is incomplete without providing at least one method
to generate {\em exactly} the distribution of a single step forward,
$(t_0,x_0,K_0)\to (t_1,z_1x_0,K_1) $,
in the primary Markovian algorithm
\begin{equation}
  d\omega(t_1,z_1x_0,K_1 | t_0,x_0,K_0)
   =\Theta(t_1-t_0)\; \Peu_{K_1 K_0}^\Theta (t_1,z_1)\; 
                      e^{-T_{K_0}(t_1,t_0)} dt_1 dz_1.
\end{equation}
The natural method of generating
the above 3-dimensional distribution (including one discrete variable)
can be read from the reorganized normalization integral
\begin{equation}
 \begin{split}  
 1 \equiv &\int\limits_{t_0}^\infty dt_1\;  \sum_{K_1}\; 
    \int\limits_0^1 dz_1\; \omega(t_1,z_1x_0,K_1 | t_0,x_0,K_0)\\
 =&\int\limits_0^1 d\big(e^{-T_{K_0}(t_1,t_0)}\big)\;
   \sum_{K_1}\; 
   \frac{\int\limits dz'\; \Peu^\Theta_{K_1K_0}(t_1,z')}%
        {\sum\limits_{X=q,g} \int\limits dz'\; \Peu^\Theta_{XK_0}(t_1,z')}
   \int\limits_0^1 dz_1\;
   \frac{\Peu^\Theta_{K_1K_0}(t_1,z_1)}{\int\limits dz'\; \Peu^\Theta_{K_1K_0}(t_1,z')}\\
 =&\int\limits_0^1 dr(t_1)\; 
   \sum_{K_1}\;    p(K_1|t_1)\;
   \int\limits_0^1 dz_1\; p(z_1|K_1,t_1)\,,
\end{split}
\end{equation}
where the two final integrals and the parton sum are each equal to $1$ separately.

One may generate the
first variable $t_1$ by inverting the cumulative distribution $r(t_1)$.
Because of the possible $t$-dependence of the coupling constant and the cut-off
parameters, this requires inverting the distribution $r(t_1)$ numerically
or preparing look-up tables for $T_K(t_1,t_0)$ form factors and their inverse,
for each parton type $K$ separately.

Knowing $t_1$,  one can generate the parton type $K_1$ according to the
probability
$\pi_{K_1K_0}$  proportional to $\int\limits dz\; \Peu^\Theta_{K_1K_0}(t_1,z)$.
Look-up tables of the $t_1$ dependent $\pi_{K_1K_0}$ branching ratios
are needed for better efficiency.

Finally, knowing $t_1$ and $K_1$ one can generate the variable $z_1$ according 
to the probability distribution proportional to 
$\Peu^\Theta_{K_1K_0}(t_1,z_1)$.
Here one can generate $z_1$ starting from some approximate distribution
and execute an internal rejection loop with the correcting weight,
about which the external part of the MC algorithm knows nothing.

As one can see, $\omega(t_1,z_1x_0,K_1 | t_0,x_0,K_0)$ can be
generated exactly. 
However, because of the need to pretabulate the form factors
and the branching probabilities there will always be some irreducible
numerical bias in the MC results. This requires some extra effort to control
quantitatively and reduce, if necessary.

\subsection{Weighted Markovian algorithm}
\label{sec:NonStandardMarkov}

The above Markovian scenario is close to what is used in the 
standard parton-shower MCs.
Here we shall describe another class of MC solutions
for Eq.~(\ref{eq:IterBasic}).
We shall stay within the class of the Markovian algorithms,
but our aim will be to use a MC implementation which allows 
for easy and quick transition to constrained Markovian algorithms.
Quite generally, in the MC algorithm described above,
there is a tendency of ``micromanaging'' the generation of the
component sub-distributions (i.e. $\omega$ distributions)
such that they are generated exactly
and there is only one extra global MC weight
of Eq.~(\ref{eq:wt_delta}).
This is an efficient method but the efficiency comes
at a price of using look-up
tables for generation of the $\omega$-distributions.

The alternative (implemented in the MC program 
{\tt EvolFMC}~\cite{EvolFMC:2005}) is to simplify intelligently
the kernels, phase space boundaries, coupling constant, etc.,
such that all component distributions in the MC algorithm
are easily generated.
The compensating weight is applied
at a later stage to exactly retrieve the original distributions
and integrals.
This also comes at a price because an extra weight will lead
to a wider weight distribution and a less efficient algorithm,
especially if one wants to turn weighted events into unweighted
ones at the  end of the MC generation.
These negative aspects can be minimized by a better choice of
approximations and the use of internal rejection loops, wherever possible.
On the positive side, there is no need to deal with 
annoying procedures of controlling quantitatively the numerical bias
due to the use of the look-up tables.
Moreover, since the approximate distributions reflect well the singularity
structure of the integrand, we have better insight into the physics
and a better chance to move away from the Markovian algorithm
(if we find it profitable for some other reasons).

Looking into the LO and NLO evolution kernels in QCD, 
one can see that they all have the following structure%
\begin{equation}
  P_{IK}(t,z)\,=\, \frac{1}{(1-z)_+}\delta_{IK} A_{KK}(t) \,+ \,
     \delta(1-z)      \delta_{IK} B_{KK}(t)
    \,+\,\frac{1}{z}                  C_{IK}(t)
                                 \,+\,D_{IK}(t,z),
\label{pik}
\end{equation}
where $D_{IK}(z)$ is completely regular.
The coefficient constants $A_{KK},B_{KK},C_{IK}$
and the coefficient functions $D_{IK}(z)$ 
can be decomposed into the LO and NLO parts:%
\begin{eqnarray}
\nonumber
A_{KK}(t) \!\!\!&=&\!\!\! \frac{\alpha_s(t)}{2\pi} A^{(0)}_{KK} 
                   +\Bigl(\frac{\alpha_s(t)}{2\pi}\Bigr)^2 A^{(1)}_{KK},
~~~~
  B_{KK}(t) \,=\, \frac{\alpha_s(t)}{2\pi} B^{(0)}_{KK} 
                   +\Bigl(\frac{\alpha_s(t)}{2\pi}\Bigr)^2 B^{(1)}_{KK}
\\\label{abcd}
\\\nonumber
  C_{IK}(t)\!\!\! &=&\!\!\! \frac{\alpha_s(t)}{2\pi} C^{(0)}_{IK} 
                   +\Bigl(\frac{\alpha_s(t)}{2\pi}\Bigr)^2 C^{(1)}_{IK},
~~~~
 D_{IK}(t,z) \,=\, \frac{\alpha_s(t)}{2\pi} D^{(0)}_{IK}(z) 
                   +\Bigl(\frac{\alpha_s(t)}{2\pi}\Bigr)^2 D^{(1)}_{IK}(z)\,.
\end{eqnarray}
Once we have made the above decomposition, we may readily
express all the form factors and constants entering into
our integrals of Eqs.~(\ref{eq:IterBasic}) and (\ref{eq:EvolIter2}).

The virtual diagonal IR-divergent elements in the kernel matrix 
and the corresponding form factor
read as follows
\begin{equation}
\begin{split}
 &\Peu^\delta_{KK}(t)= 
         2
         \left( A_{KK}(t) \ln\frac{1}{\epsilon(t)} -B_{KK}(t)\right),\\
 & \Phi_K(t,t_0)=
      \int\limits_{t_0}^{t} dt'\; \Peu^\delta_{KK}(t')
   = \int\limits_{t_0}^{t} dt'\;
      2 \left( A_{KK}(t') \ln\frac{1}{\epsilon(t')} -B_{KK}(t')\right),\\
\end{split}
\end{equation}
The integrated real-emission off-diagonal elements
needed to generate the parton type in the Markovian MC is now
\begin{equation}
 \pi_{IK}(t) =\int\limits_{0}^{1} dz\;\Peu^\Theta_{IK}(t,z)
             = 
               2 \bigg[
	       \delta_{IK} A_{KK}(t) \ln\frac{1}{\epsilon(t)}
	      +            C_{IK}(t) \ln\frac{1}{\epsilon'}
	      +\int\limits_0^1 dz\; D_{IK}(t,z) 
               \bigg]\,.
\end{equation}
The real-emission form factors $T_K,K=q,g$, are given by
\begin{equation}
\begin{split}
  T_K(t,t_0)
  &=\int\limits_{t_0}^{t} dt'\; \sum_{X} \pi_{XK}(t')\\
  &=\int\limits_{t_0}^{t} dt'\;
      2\bigg[
     A_{KK}(t') \ln\frac{1}{\epsilon(t')}
    +\sum_{X} C_{XK}(t') \ln\frac{1}{\epsilon'}
    +\sum_{X}\int\limits_0^1 dz\; D_{XK}(t',z) 
    \bigg].
\end{split}
\end{equation}
Finally, the form factors $\Delta_K$, needed in the final MC weight
for the correct overall normalization, read
\begin{equation}
\begin{split}
  \Delta_K(t,t_0)
  &= \int\limits_{t_0}^{t} dt' \sum_{X} \int\limits dz\; \Peu_{XK}(t')
   =\int\limits_{t_0}^{t} dt'
    \bigg\{ -\Peu^\delta_{KK}(t')
            +\sum_{X} \pi_{XK}(t')\bigg\}\\
  &= \int\limits_{t_0}^{t} dt'\;
     2 \bigg[
     B_{KK}(t')
    +\sum_{X} C_{XK}(t') \ln\frac{1}{\epsilon'}
    +\sum_{X}\int\limits_0^1 dz\; D_{XK}(t',z) 
    \bigg].
\end{split}
\end{equation}

Having expressed all the elements in Eq.~(\ref{eq:MarkovianStandard})
of the standard Markovian algorithm, let us construct
an alternative MC Markovian scenario
starting from Eq.~(\ref{eq:IterBasic}) (before Markovianization).
First, we simplify the kernel matrix elements 
\begin{equation}
\label{eq:KernSimp1}
\begin{split}
  \Peu^\Theta_{IK}(t,z)
 &\,\to\, \hat{\Peu}^\Theta_{IK}(t_0,z)
   = \Theta(z-\epsilon')\,\Theta(1-z-\hat\epsilon)\,
       \frac{\alpha_s(t_0)}{\pi} \\
 & \,\times\,\bigg\{\frac{1}{1-z} \,\delta_{IK} A^{(0)}_{KK}
           +\frac{1}{z}  \,                C^{(0)}_{IK}
	                                +\hat{D}_{IK}\bigg\},
\end{split}
\end{equation}
where 
$D$ is replaced by the constant $\hat{D}$,
which is chosen to be zero when $A^{(0)},B^{(0)}$ are nonzero
or equal the maximum (positive) value of $D^{(0)}_{IK}(z)$; 
see Appendix~\ref{sec:kernels1}.
The above simplification is of course compensated by the MC weight
\begin{equation}
  w_P= \prod_{i=1}^n \frac{      \Peu^\Theta_{K_iK_{i-1}}(t_i,z_i)}%
                          {\hat{\Peu^\Theta}_{K_iK_{i-1}}(t_0,z_i)}
 .
\end{equation}
Let us remark that the replacement $\alpha_s(t_i)\to\alpha_s(t_0)$
of the running coupling which stands in front of the LO kernel
might cause poor overall MC efficiency. 
This problem is addressed separately in the next section.

The above reorganization leads us to the following new formula 
\begin{equation}
  \label{eq:Iter3}
  \begin{split}
  D_K&(t,x) = e^{-\Phi_K(t,t_0)} D_K(t_0,x)
  +\sum_{n=1}^\infty \;
   \sum_{K_0,\ldots,K_{n-1}}
      \int\limits_0^1 dx_0\;
      \prod_{i=1}^n \bigg[ \int\limits_{t_0}^t dt_i\; \Theta(t_i-t_{i-1}) \int\limits_0^1 dz_i\bigg]
\\&~~\times
      e^{-\Phi_K(t,t_n)}
      \prod_{i=1}^n 
          \bigg[ \hat\Peu_{K_iK_{i-1}}^\Theta (t_0,z_i) 
                 e^{-\Phi_{K_{i-1}}(t_i,t_{i-1})} \bigg]
      D_{K_0}(t_0,x_0) \delta\big(x- x_0\prod_{i=1}^n z_i \big) w_P,
  \end{split}
\end{equation}
completely equivalent to Eq.~(\ref{eq:IterBasic}).
Markovianization is now done for the variant of the above
formula in which $w_P$ is neglected.
We define a new transition probability as follows
\begin{equation}
  \begin{split}  
    &\hat\omega(t_i,x_i,K_i| t_{i-1},x_{i-1},K_{i-1})
    \equiv  \Theta(t_i-t_{i-1})\;
    \hat\Peu_{K_iK_{i-1}}^\Theta (t_0,x_i/x_{i-1})\;
    e^{-\hat{T}_{K_{i-1}}(t_i,t_{i-1})},\\
    &\int\limits_{t_{i-1}}^\infty dt_i\; \int\limits_0^1 dz_i \sum_{K_i}
    \hat\omega(t_i,x_i,K_i| t_{i-1},x_{i-1},K_{i-1})\equiv 1,
  \end{split}
\end{equation}
where
\begin{equation}
\begin{split}
 \hat{T}_K(t_{i},t_{i-1})
  &= \int\limits_{t_{i-1}}^{t_{i}} dt'\; 
             \int\limits_{0}^{1} dz\;
             \sum_J \hat\Peu^\Theta_{JK}(t',z)
\\
  &=  (t_{i}-t_{i-1}) \frac{\alpha_s(t_0)}{\pi}
     \bigg[
                     A^{(0)}_{KK} \ln\frac{1}{\hat{\epsilon}}
      +\sum_{X}      C^{(0)}_{XK} \ln\frac{1}{{\epsilon}'}
      +\sum_{X} \hat{D}^{(0)}_{XK} 
     \bigg]
\\
  &= \int\limits_{t_{i-1}}^{t_{i}} dt'\; \sum_{X} \hat\pi_{XK}
   = (t_{i}-t_{i-1}) \sum_{X} \hat\pi_{XK} = (t_{i}-t_{i-1})\; R_K\,,
\\
\end{split}
\label{hatt}
\end{equation}
and the probability rate of the parton transition $K\to I$ is now constant
\begin{equation}
\hat\pi_{IK} 
  =\int\limits_{0}^{1} dz\;\hat\Peu^\Theta_{IK}(t_0,z)
             =\frac{\alpha_s(t_0)}{\pi} \bigg[
	       \delta_{IK} A^{(0)}_{KK} \ln\frac{1}{\hat{\epsilon}}
	      +            C^{(0)}_{IK} \ln\frac{1}{{\epsilon}'}
	      +\hat{D}_{IK}
               \bigg],
\label{hatpi}
\end{equation}
independent of $t$; see Appendix~\ref{sec:kernels1} for explicit formulae. 
Summarizing, the transition probability
\begin{equation}
  \hat\omega(t_i,x_i,K_i| t_{i-1},x_{i-1},K_{i-1})
    \equiv  \Theta(t_i-t_{i-1})\;
    \hat\Peu_{K_iK_{i-1}}^\Theta (t_0,x_i/x_{i-1})\;
    e^{-(t_i-t_{i-1}) R_{K_{i-1}}}
\end{equation}
is now such a simple function that can be generated using elementary MC methods,
without any pretabulation.
The last thing necessary, as usual, for the Markovianization is
introduction of the ``spill-over'' variable. This is done with the
help of the identity 
\begin{equation}
  e^{-\Phi_{K_{n}}(t,t_{n})}=
  e^{\hat\Delta_{K_{n}}(t,t_{n})}
  \int\limits_{t}^\infty dt_{n+1}\; \int\limits_0^1 dz_{n+1} \sum_{K_{n+1}}
  \hat\omega( t_{n+1},x_{n+1},K_{n+1}| t_n,x_n,K_n)\,,
\end{equation}
where
\begin{equation}
  \hat\Delta_{K}(t_{i},t_{i-1}) 
          = \hat{T}_{K}(t_{i},t_{i-1})-\Phi_{K}(t_{i},t_{i-1})
          = (t_{i}-t_{i-1}) R_{K} -\Phi_{K}(t_{i},t_{i-1})\,.
\end{equation}
Let us stress that now, contrary to the previous standard Markovian scenario,
$\hat\Delta$ has an explicit residual dependence on the IR cut $\hat\epsilon$
which is necessary to cancel {\em exactly} the analogous
dependence of the average weight $w_P$
(similarly as in a typical MC algorithm for QED exponentiation).

Summarizing all the above discussion, we transform Eq.~(\ref{eq:Iter3})
in a new equivalent form
\begin{equation}
  \label{eq:Markovian2}
  \begin{split}
  D_K(t,x) &= 
  e^{\hat\Delta_{K}(t,t_{0})}
  \int\limits_{t_{1}>t} dt_{1} dz_{1} \sum_{K_{1}}
  \hat\omega( t_{1},x_{1},K_{1}| t_0,x,K)\;\;
  D_K(t_0,x)
\\&
  +\sum_{n=1}^\infty \;\;
  \int\limits_0^1 dx_0\;
  \int\limits_{t_{n+1}>t} dt_{n+1} dz_{n+1} \sum_{K_{n+1}}\;\;
  \sum_{K_0,\ldots,K_{n-1}}\;\;
      \prod_{i=1}^n \int\limits_{t_i<t} dt_idz_i \;
\\&~~~~~~~~~\times
   \hat\omega( t_{n+1},x_{n+1},K_{n+1}| t_n,x_n,K_n)\;
   \prod_{i=1}^n 
      \hat\omega(t_i,x_i,K_i| t_{i-1},x_{i-1},K_{i-1})
\\&~~~~~~~~~\times
      \delta\big(x- x_0\prod_{i=1}^n z_i \big)\;
      D_{K_0}(t_0,x_0)\; w_P w_\Delta \,,
  \end{split}
\end{equation}
where
\begin{equation}
 \label{eq:wt_delta2}
  w_\Delta=e^{\hat\Delta_{K_{n}}(t,t_{n})} 
      \prod_{i=1}^n e^{\hat\Delta_{K_{i-1}}(t_i,t_{i-1})}\,.
\end{equation}
In the MC generation we proceed as before. Neglecting the weight $w=w_P w_\Delta$,
we generate primary MC events using the Markovian
algorithm with the simplified transition probability $\hat\omega$.
The original distributions and integrals
are recovered by applying the correction weight $w$.
As already stressed, the MC efficiency will be worse than in the previous case, 
but the whole MC program is now much simpler and most likely
provides better control of the technical precision.

Let us note that we shall still need a precise $1$-dimensional
pretabulation of all the form factors $\Phi_K(t,t_0)$, 
entering into $w_\Delta$ through $\hat\Delta_K$.

\subsection{Importance sampling for running $\alpha_s(t)$}

In the above we took into account the $t$-dependence $(t=\ln Q)$,
that is running, of the strong coupling constant $\alpha_s(t)$
by reweighting MC events. This is very inefficient
and it is rather easy to introduce the relevant $t$-dependence
of $\alpha_s(t)$ at least at the one-loop level
\begin{equation}
  \label{eq:alfS1}
  \alpha^{(0)}_s(t)= \frac{4\pi}{ \beta_0 (2t-2\ln\Lambda_0)}
\end{equation}
already in the underlying MC distributions.

One can see that in the $t$-integration in Eq.~(\ref{eq:Iter3}) we have
effectively
\begin{equation}
\int\limits dt_i\; \alpha^{(0)}_s(t_i) 
     = \int\limits dt_i\; \frac{2\pi}{\beta_0 (t_i-\ln\Lambda_0)}
     = \frac{2\pi}{\beta_0} \int\limits d \ln(t_i-\ln\Lambda_0).
\end{equation}
It is therefore natural to introduce the variable $\tau_i$,
\begin{equation}
  \begin{split}
  &\tau_i = \ln(t_i-\ln\Lambda_0),\quad t_i= \ln\Lambda_0 +\exp(\tau_i),\\
  &dt_i = (t_i-\ln\Lambda_0) d\tau_i = e^{\tau_i} d\tau_i,
  \end{split}
\end{equation}
instead of%
\footnote{Of course, we are aware of a possibility of
  introducing $\tau$ as an evolution ``time'' in the original
  differential equation from the very beginning. We proceed this way
  in order to get more insight into various versions of the MC algorithm.}
$t_i$.
This change of variables will lead to the Jacobian factor 
$e^{\tau_i}=t_i-\ln\Lambda_0$ in the integrand,
which will cancel the unwanted factor $e^{-\tau_i}=(t-\ln\Lambda_0)^{-1}$ 
present in $\alpha_s(t)$ in the MC weight .
The two-loop and more complicated contributions to $\alpha_s(t)$
may still be added by reweighting events, without spoiling much
the efficiency.

We start again from Eq.~(\ref{eq:IterBasic}) (before Markovianization).
The kernel matrix elements are now simplified more ``gently''
with respect to Eq.~(\ref{eq:KernSimp1})
\begin{equation}
\begin{split}
  \Peu^\Theta_{IK}(t,z)
 &\,\to\, \bar{\Peu}^\Theta_{IK}(t,z)
   = \Theta(z-\epsilon')\,\Theta(1-z-\bar\epsilon)\,
       \frac{\alpha^{(0)}_s(t)}{\pi} \\
 &\, \times\,\bigg\{\frac{1}{(1-z)_+}\, \delta_{IK} A^{(0)}_{KK}
           \,+\,\frac{1}{z}\,                  C^{(0)}_{IK}
	                               \, +\,\bar{D}_{IK}\bigg\},
\end{split}
\end{equation}
where $\bar{D}_{IK}=\hat{D}_{IK}$ and 
$\epsilon(t)\to\bar\epsilon$.
The new compensating MC weight is
\begin{equation}
  \bar w_P= \prod_{i=1}^n \frac{  \Peu^\Theta_{K_iK_{i-1}}(t_i,z_i)}%
                           {  \bar\Peu^\Theta_{K_iK_{i-1}}(t_i,z_i)}
.
\end{equation}

The above reorganization leads us to the following new formula, 
completely equivalent to Eq.~(\ref{eq:IterBasic}),
\begin{equation}
  \label{eq:Iter5}
  \begin{split}
  D_K&(\tau,x) = e^{-\Phi_K(\tau,\tau_0)} D_K(\tau_0,x)
  +\sum_{n=1}^\infty \;
   \sum_{K_0,\ldots,K_{n-1}}
   \prod_{i=1}^n \bigg[ \int\limits_{\tau_0}^\tau d\tau_i\; 
     \Theta(\tau_i-\tau_{i-1}) \int\limits_0^1 dz_i\bigg]
\\&\times
      e^{-\Phi_K(\tau,\tau_n)}
      \int\limits_0^1 dx_0
      \prod_{i=1}^n 
          \bigg[ e^{\tau_i} \bar\Peu_{K_iK_{i-1}}^\Theta (\tau_i,z_i) 
                 e^{-\Phi_{K_{i-1}}(\tau_i,\tau_{i-1})} \bigg]
      D_{K_0}(\tau_0,x_0) \delta\big(x- x_0\prod_{i=1}^n z_i \big) \bar{w}_P.
  \end{split}
\end{equation}
In the following we shall use a new function
\begin{equation}
  \bar{\bar\Peu}_{K_iK_{i-1}}^\Theta (z_i)=
  e^{\tau_i}\; \bar\Peu_{K_iK_{i-1}}^\Theta (\tau_i,z_i)\,,
\end{equation}
because it {\em does not depend} on $\tau_i$ anymore.
This is the whole point of the $t_i \to \tau_i$ change of variables.

Omitting $\bar{w}_P$, we proceed to Markovianization 
following the example of the previous section
\begin{equation}
  \begin{split}  
    &\bar\omega(\tau_i,x_i,K_i| \tau_{i-1},x_{i-1},K_{i-1})
    \equiv  \Theta(\tau_i-\tau_{i-1})\;
    \bar{\bar\Peu}_{K_iK_{i-1}}^\Theta (x_i/x_{i-1})\;
    e^{-\bar{T}_{K_{i-1}}(\tau_i,\tau_{i-1})},\\
    &\int\limits_{\tau_{i-1}}^\infty d\tau_i\; \int\limits_0^1 dz_i \sum_{K_i}
    \bar\omega(\tau_i,x_i,K_i| \tau_{i-1},x_{i-1},K_{i-1})\equiv 1\,,
  \end{split}
\end{equation}
where
\begin{equation}
\begin{split}
 \bar{T}_K(\tau_{i},\tau_{i-1})
  &= \int\limits_{\tau_{i-1}}^{\tau_{i}} d\tau'\; 
             \int\limits_{0}^{1} dz\;
             \sum_J \bar{\bar\Peu}^\Theta_{JK}(z)
\\
  &=  (\tau_{i}-\tau_{i-1})
     \frac{2}{\beta_0}
     \bigg[
                     A^{(0)}_{KK} \ln\frac{1}{\bar{\epsilon}}
      +\sum_{X}      C^{(0)}_{XK} \ln\frac{1}{{\epsilon}'}
      +\sum_{X} \bar{D}^{(0)}_{XK} 
     \bigg]
\\
  &= (\tau_{i}-\tau_{i-1}) \bar R_K\,, 
\\
  \bar R_K\ &= \sum_{X} \bar\pi_{XK}
\end{split}
\end{equation}
and the probability rate of the parton transition $K\to I$ is now a constant
\begin{equation}
\bar\pi_{IK} 
  =\int\limits_{0}^{1} dz\;
       \bar{\bar\Peu}^\Theta_{IK}(z)
             =\frac{2}{\beta_0} \bigg[
	       \delta_{IK} A^{(0)}_{KK} \ln\frac{1}{\bar\epsilon}
	      +            C^{(0)}_{IK} \ln\frac{1}{\epsilon'}
	      +\bar{D}_{IK}
               \bigg],
\end{equation}
again independent of $t$. 
The final transition probability to be generated in each step
of the Markovian algorithm reads
\begin{equation}
  \bar\omega(\tau_i,x_i,K_i| \tau_{i-1},x_{i-1},K_{i-1})
    \equiv  \Theta(\tau_i-\tau_{i-1})\;
    \bar{\bar\Peu}_{K_iK_{i-1}}^\Theta (x_i/x_{i-1})\;
    e^{-(\tau_i-\tau_{i-1}) \bar R_{K_{i-1}}},
\end{equation}
where
\begin{equation}
\begin{split}
   \bar{\bar\Peu}^\Theta_{IK}(z)
    &= \Theta(z-\epsilon')\,\Theta(1-z-\bar\epsilon)\,
    \frac{2}{\beta_0}\\
   &\times\bigg\{
           \frac{1}{1-z}\,   \delta_{IK} A^{(0)}_{KK}
           +\frac{1}{z}\,                  C^{(0)}_{IK}
	                                +\bar{D}_{IK}\bigg\}
\\
\end{split}
\end{equation}
is again a simple function which can be generated, without
any pretabulation, using elementary MC methods.
The overall recipe, as compared with the previous MC algorithm, is
to replace: $t_i\to \tau_i$ and 
$\alpha_s(t_0)/\pi\to 2/\beta_0$
in the generation of the primary MC distribution $\bar\omega$,
before applying $\bar{w}_P$.

Inevitably, to complete the Markovianization,
the integral over a ``spill-over'' variable $\tau_{n+1}$
is added with the usual identity
\begin{equation}
  e^{-\Phi_{K_{n}}(\tau,\tau_{n})}=
  e^{\bar\Delta_{K_{n}}(\tau,\tau_{n})}
  \int\limits_{\tau}^\infty d\tau_{n+1}\; \int\limits_0^1 dz_{n+1} \sum_{K_{n+1}}
  \bar\omega( \tau_{n+1},x_{n+1},K_{n+1}| \tau_n,x_n,K_n)\,,
\end{equation}
where
\begin{equation}
  \bar\Delta_{K}(\tau_{i},\tau_{i-1}) 
          = \bar{T}_{K}(t_{i},\tau_{i-1})-\Phi_{K}(\tau_{i},\tau_{i-1})
          = (\tau_{i}-\tau_{i-1}) \bar R_{K}  -\Phi_{K}(\tau_{i},\tau_{i-1})\,.
\end{equation}

The final formula, equivalent to original Eq.~(\ref{eq:Iter3}),
for this MC scenario with the importance sampling for the
running $\alpha_s$ reads as follows
\begin{equation}
  \label{eq:Markovian3}
  \begin{split}
  D_K(\tau,x) &= 
  e^{\bar\Delta_{K}(\tau,\tau_{0})}
  \int\limits_{\tau_{1}>\tau} d\tau_{1} dz_{1} \sum_{K_{1}}
  \bar\omega( \tau_{1},x_{1},K_{1}| \tau_0,x,K)\;\;
  D_K(\tau_0,x)
\\&
  +\sum_{n=1}^\infty \;\;
      \int\limits_0^1 dx_0\;
  \int\limits_{\tau_{n+1}>\tau} d\tau_{n+1} dz_{n+1} \sum_{K_{n+1}}\;\;
  \sum_{K_0,\ldots,K_{n-1}}\;\;
      \prod_{i=1}^n \int\limits_{\tau_i<\tau} d\tau_idz_i \;
\\&~~~~~~~~~\times
   \bar\omega( \tau_{n+1},x_{n+1},K_{n+1}| \tau_n,x_n,K_n)\;
   \prod_{i=1}^n 
      \bar\omega(\tau_i,x_i,K_i| \tau_{i-1},x_{i-1},K_{i-1})
\\&~~~~~~~~~\times
      \delta\big(x- x_0\prod_{i=1}^n z_i \big)\;
      D_{K_0}(\tau_0,x_0)\; \bar w_P \bar w_\Delta \,,
  \end{split}
\end{equation}
where
\begin{equation}
 \label{eq:wt_delta3}
  \bar w_\Delta=e^{\bar\Delta_{K_{n}}(\tau,\tau_{n})} 
      \prod_{i=1}^n e^{\bar\Delta_{K_{i-1}}(\tau_i,\tau_{i-1})}\,.
\end{equation}
The above formula looks almost identical to Eq.~(\ref{eq:Markovian2}).

\section{Markovian MC for parton-momentum distributions}
\label{sec-markowian-x}

The factor $1/z$ in the brems\-strahlung kernels causes
a significant loss of MC efficiency 
due to $\exp(\Delta_K)$ which contains uncompensated $\ln(\epsilon')$.
We can get rid of this annoying phenomenon by switching
to the $xD(x)$ which evolve with the kernels $zP(z)$.
The reason for improvement is that kernels $zP(z)$
fulfill  the momentum-conservation sum rules.
The evolution equations for $xD(x)$ read%
\begin{equation}
  \label{eq:Evolu2}
  \partial_t\, xD_K(t,x)
   = \sum_J \int\limits_x^1 \frac{d z}{z}\; z\Peu_{KJ}(t,z)\; 
    \frac{x}{z} D_J\Big(t,\frac{x}{z} \Big)\,.
\end{equation}

The iterative solution can be obtained from the above formulae,  
or equivalently by multiplying both sides of Eq.~(\ref{eq:IterBasic}) by $x$,
\begin{equation}
  \label{eq:Iter6}
  \begin{split}
  xD_K&(t,x) = e^{-\Phi_K(t,t_0)} xD_K(t_0,x)
  +\sum_{n=1}^\infty \;
  \int\limits_0^1 dx_0\;
   \sum_{K_0,\ldots,K_{n-1}}
      \prod_{i=1}^n \bigg[ \int\limits_{t_0}^t dt_i\; \Theta(t_i-t_{i-1}) \int\limits_0^1 dz_i\bigg]
\\&~~~\times
      e^{-\Phi_K(t,t_n)}
      \prod_{i=1}^n 
          \bigg[ 
                 z_i\Peu_{K_i K_{i-1}}^\Theta (t_i,z_i) 
                 e^{-\Phi_{K_{i-1}}(t_i,t_{i-1})} \bigg]
      x_0 D_{K_0}(t_0,x_0)\, \delta\big(x- x_0\prod_{i=1}^n z_i \big),
  \end{split}
\end{equation}
where $K\equiv K_n$.
It was essential to exploit the condition $x= x_0\prod_{i=1}^n z_i$
imposed by the overall $\delta$-functions%
\footnote{This way we could introduce any power of $x$, say $x^\alpha$,
in front of $D(x)$ and $z^\alpha$ in the kernels.}.
We also feel free to introduce such an overall factor $x$ in the $xD(x)$,
because our ultimate aim is to use a constrained Markovian algorithm, 
hence such a factor
will be dealt in the MC separately and independently with other
dedicated MC methods.
At the technical level, we may multiply both sides of the above equation
by $1/x$ and absorb $1/x$ in the MC weight, pretending that we generate
$D(x)$ distribution as before; in such a case the fluctuations of the weight 
in the histograms of $x$ will change but the distribution of $x$
will be the same.
The main change will be in the probability distribution
$\omega$ for the forward leap in the Markovian random walk.

Before we enter into details of the Markovian MC, let us introduce the
evolution 
variable $\tau$, similarly as in the previous section
\begin{equation}
\label{eq:tau1}
\tau \equiv \frac{1}{\alpha_s(t_A)} \int\limits_{t_A}^{t} dt_1\; \alpha_s(t_1),\quad
\frac{\partial t}{\partial\tau}= \frac{\alpha_s(t_A)}{\alpha_s(t)}.
\end{equation}
In the above transformation we may use various choices of $t_A$
and of $\alpha_s(t)$. For instance, we may employ the same $\alpha_s(t)$
as in the evolution equations (LO or NLO) or we may stay
with the one-loop LO: 
$\alpha_s^{(0)}(t)= 2\pi/( \beta_0 (t-\ln\Lambda_0))$.
In the latter case, with $t_A$ chosen such that
$\alpha_s^{(0)}(t_A)=2\pi/\beta_0$ 
(e.g. $t_A-\ln\Lambda_0=1$, $t_A=\ln(e\Lambda_0)$),
we recover the definition $\tau=\ln(t-\ln\Lambda_0)$ of the previous section.
Let us adjust $t_A=t_0$ to the starting point of the evolution
and use $\alpha_s^{(0)}(t)$ in the definition of $\tau$.
With such a choice we have
\begin{equation}
  \label{eq:Iter7}
  \begin{split}
  xD_K&(\tau,x) = e^{-\Phi_K(\tau,\tau_0)} xD_K(\tau_0,x)
  +\sum_{n=1}^\infty \;
  \int\limits_0^1 dx_0
   \sum_{K_0,\ldots,K_{n-1}}
   \prod_{i=1}^n \bigg[ \int\limits_{\tau_0}^\tau d\tau_i\; 
     \Theta(\tau_i-\tau_{i-1}) \int\limits_0^1 dz_i\bigg]
\\&~~\times
      e^{-\Phi_K(\tau,\tau_n)}
      \prod_{i=1}^n 
          \bigg[ \Pcal_{K_i K_{i-1}}^\Theta (\tau_i,z_i) 
                 e^{-\Phi_{K_{i-1}}(\tau_i,\tau_{i-1})} \bigg]
      x_0D_{K_0}(\tau_0,x_0)\, \delta\big(x- x_0\prod_{i=1}^n z_i \big),
  \end{split}
\end{equation}
where $K\equiv K_n$, and
\begin{equation}
  \Pcal_{K_i K_{i-1}}^\Theta (\tau_i,z_i)=
   \frac{\alpha_s^{(0)}(t_0)}{\alpha_s^{(0)}(t_i)}\;
    z_i\Peu_{K_i K_{i-1}}^\Theta (\tau_i,z_i),
\end{equation}
which depends on $\tau_i$ very weakly.
In the LO case it is completely independent of $\tau_i$. 

In the general (NLO) case
we may decompose the evolution kernels multiplied by $z$ as follows
\begin{equation}
\label{zpeu}
 \frac{1}{2}
   z\Peu_{IK}(t,z)\equiv zP_{IK}(t,z)= \frac{1}{(1-z)_+}\delta_{IK} A_{KK}(t)+
             \delta(1-z) \delta_{IK}      B_{KK}(t)
	                                 +F_{IK}(t,z),
\end{equation}
where the finite part $F$ can be expressed in terms of the previously
defined functions $A,\, C$ and $D$ as follows
\begin{equation}
  F_{IK}(t,z) = zD_{IK}(t,z)+C_{IK}(t)-\delta_{IK} A_{KK}(t).
\end{equation}

In the MC we replace, as before, the full kernels for the 
real emission with the LO approximation
with the constant IR regulator $\bar\epsilon\leq \epsilon(t)$:
\begin{equation}
\label{barpcal}
\begin{split}
 &\Pcal^\Theta_{IK}(\tau,z)
  \to \bar{\Pcal}^\Theta_{IK}(\tau_0,z)
   = \Theta(1-z-\bar\epsilon)
       \frac{\alpha_s^{(0)}(t_0)}{\pi} z P^{(0)}_{IK}(z),\\ \\
 & z P^{(0)}_{IK}(z)=\frac{1}{(1-z)_+} \delta_{IK} A^{(0)}_{KK}+
            \delta(1-z)    \delta_{IK} B^{(0)}_{KK}
	                              +F^{(0)}_{IK}(z).
\end{split}
\end{equation}
The approximate kernels do not depend on $\tau$.
The corresponding compensating weight is now
\begin{equation}
  \label{eq:wtZ}
  \bar w_P= \prod_{i=1}^n \frac{  \Pcal^\Theta_{K_i K_{i-1}}(\tau_i,z_i)}%
                           {  \bar\Pcal^\Theta_{K_i K_{i-1}}(\tau_0,z_i)}
.
\end{equation}

The probability of the forward Markovian leap reads now as follows
\begin{equation}
  \begin{split}  
    &\bar\omega(\tau_i,x_i,K_i| \tau_{i-1},x_{i-1},K_{i-1})
    \equiv  \Theta(\tau_i-\tau_{i-1})\;
    \bar\Pcal_{K_i K_{i-1}}^\Theta (\tau_0,x_i/x_{i-1})\;
    e^{-\bar{T}_{K_{i-1}}(\tau_i,\tau_{i-1})},\\ 
    &\int\limits_{\tau_{i-1}}^\infty d\tau_i\; \int\limits_0^1 dz_i \sum_{K_i}
    \bar\omega(\tau_i,x_i,K_i| \tau_{i-1},x_{i-1},K_{i-1})\equiv 1\,,
    \;\; z_i=x_i/x_{i-1},
  \end{split}
\end{equation}
where the new real-emission form factor is defined as follows
\begin{equation}
\begin{split}
 \bar{T}_K(\tau_{i},\tau_{i-1})
  &= \int\limits_{\tau_{i-1}}^{\tau_{i}} d\tau'\; 
             \int\limits_{0}^{1} dz\;
             \sum_J \bar\Pcal^\Theta_{JK}(\tau_0,z)
\\
  &=  (\tau_{i}-\tau_{i-1})
     \frac{\alpha_s^{(0)}(t_0)}{\pi}
     \bigg[
                     A^{(0)}_{KK} \ln\frac{1}{\bar\epsilon}
      +\sum_{J} \int\limits_0^{1} F^{(0)}_{JK}(z) dz
     \bigg]
\\
  &= (\tau_{i}-\tau_{i-1}) \sum_{J} \bar \pi_{JK} 
   = (\tau_{i}-\tau_{i-1})\; \bar R_K.
\\
\end{split}
\end{equation}
The rate of the parton transition $K\to I$ is now
\begin{equation}
\label{barpi}
\bar \pi_{IK} 
  =\int\limits_{0}^{1} dz\;
       \bar\Pcal^\Theta_{IK}(\tau_0,z)
             =\frac{\alpha_s^{(0)}(t_0)}{\pi} \bigg[
	       \delta_{IK} A^{(0)}_{KK} \ln\frac{1}{\bar\epsilon}
	      +\int\limits_0^{1} F^{(0)}_{IK}(z) dz
               \bigg]\,.
\end{equation}

On the other hand, the exact virtual form factor is
\begin{equation}
\label{eq:PhiFF}
  \Phi_K(\tau,\tau_0)
   =\int\limits_{\tau_0}^{\tau} d\tau'\;
   \frac{\alpha_s^{(0)}(t_0)}{\alpha_s^{(0)}(t')}\;
     2
     \left[ A_{KK}(\tau') \ln\frac{1}{\epsilon(\tau')} -B_{KK}(\tau')\right]\,.
\end{equation}
In the LO, for the one-loop $\alpha_s^{(0)}$ and 
if, in addition, we choose $\epsilon(\tau)=\epsilon=const$, 
then the virtual form factor becomes simply
\begin{equation}
  \Phi_K(\tau,\tau_0)=
    (\tau-\tau_0)
     \frac{\alpha_s^{(0)}(t_0)}{\pi}
     \left( A_{KK}^{(0)} \ln\frac{1}{\epsilon} -B_{KK}^{(0)}\right)\,.
\end{equation}

Summarizing, the final transition probability to be generated in each step
of the Markovian algorithm reads
\begin{equation}
  \bar\omega(\tau_i,x_i,K_i| \tau_{i-1},x_{i-1},K_{i-1})
    \equiv  \Theta(\tau_i-\tau_{i-1})\;
    \bar\Pcal_{K_i K_{i-1}}^\Theta (\tau_0, x_i/x_{i-1})\;
    e^{-(\tau_i-\tau_{i-1}) \bar R_{K_{i-1}}}\,,
\end{equation}
where
\begin{equation}
   \bar\Pcal^\Theta_{IK}(\tau_0,z)
    = \Theta(1-z-\bar\epsilon)
    \frac{\alpha_s^{(0)}(t_0)}{\pi}
    \bigg\{
            \frac{1}{1-z} \delta_{IK} A^{(0)}_{KK}
	    +F^{(0)}_{IK}(z)\bigg\}\,.
\end{equation}

To complete the Markovianization,
the integral over the ``spill-over'' variable $\tau_{n+1}$
is added with the help of the usual identity
\begin{equation}
  e^{-\Phi_{K_{n}}(\tau,\tau_{n})}=
  e^{\bar\Delta_{K_{n}}(\tau,\tau_{n})}
  \int\limits_{\tau}^\infty d\tau_{n+1}\; \int\limits_0^1 dz_{n+1} \sum_{K_{n+1}}
  \bar\omega( \tau_{n+1},x_{n+1},K_{n+1}| \tau_n,x_n,K_n)\,,
\end{equation}
where $z_{n+1}=x_{n+1}/x_n$, and
\begin{equation}
  \bar\Delta_{K}(\tau_{i},\tau_{i-1}) 
          = \bar{T}_{K}(\tau_{i},\tau_{i-1})-\Phi_{K}(\tau_{i},\tau_{i-1})
          = (\tau_{i}-\tau_{i-1}) \bar R_{K}  -\Phi_{K}(\tau_{i},\tau_{i-1}).
\end{equation} 
The advantage of the method outlined in this section is that
at the LO level we obtain for $\epsilon=\bar\epsilon$
\begin{equation}
  \bar\Delta_{K}=0
\end{equation} 
due to the fact that the kernels obey the momentum sum rules.
In most renormalization schemes (e.g. $\overline{MS}$)
this will be also valid at the NLO level%
  \footnote{A small non-zero value of $\bar\Delta_{k}$ may be present
            for technical reasons, that is, if we use
            slightly simplified kernels at the low MC generation level.}.

The final formula for this MC scenario with the importance sampling for the
running $\alpha_s$ reads
\begin{equation}
  \label{eq:Markovian4}
  \begin{split}
  xD_K(\tau,x) &= 
  e^{\bar\Delta_{K}(\tau,\tau_{0})}
  \int\limits_{\tau_{1}>\tau} d\tau_{1} dz_{1} \sum_{K_{1}}
  \bar\omega( \tau_{1},z_1x,K_1| \tau_0,x,K)\;\;
  xD_K(\tau_0,x)
\\&
  +\sum_{n=1}^\infty \;\;
  \int\limits_0^1 dx_0\;
  \int\limits_{\tau_{n+1}>\tau} d\tau_{n+1} dz_{n+1} \sum_{K_{n+1}}\;\;
  \sum_{K_0,\ldots,K_{n-1}}\;\;
      \prod_{i=1}^n \int\limits_{\tau_i<\tau}^t d\tau_idz_i \;
\\&~~~~~~~~~\times
   \bar\omega( \tau_{n+1},x_{n+1},K_{n+1}| \tau_n,x_n,K_n)\;
   \prod_{i=1}^n 
      \bar\omega(\tau_i,x_i,K_i| \tau_{i-1},x_{i-1},K_{i-1})
\\&~~~~~~~~~\times
      \delta\big(x- x_0\prod_{i=1}^n z_i \big)\;
      x_0 D_{K_0}(\tau_0,x_0)\; \bar w_P \bar w_\Delta.
  \end{split}
\end{equation}
where $z_{i}=x_{i}/x_{i-1}$, $K\equiv K_{n}$ and
\begin{equation}
 \label{eq:wt_delta4}
  \bar w_\Delta=e^{\bar\Delta_{K_{n}}(\tau,\tau_{n})} 
      \prod_{i=1}^n e^{\bar\Delta_{K_{i-1}}(\tau_i,\tau_{i-1})}\,.
\end{equation}

\section{Numerical tests}
\label{sec:num-tests}

We have performed comparisons of the MC solution of the DGLAP
evolution equations implemented the program {\tt EvolFMC}~\cite{EvolFMC:2005}
with another solution provided by the non-MC program {\tt QCDnum16}~\cite{qcdnum16}.
In both cases we have evolved singlet PDF for gluons
and three doublets of massless quarks from $Q_0=1\,$GeV
to $Q=10,100,1000\,$GeV.
The comparisons have been done both for the LO and the NLO evolution,
including the running $\alpha_s$ in the corresponding approximation.

In our test, we have used the following parameterization of the starting parton 
distributions in the proton at $Q_0=1\,$GeV:
\begin{equation}
  \begin{split}
    xD_G(x)        &= 1.9083594473\cdot x^{-0.2}(1-x)^{5.0},\\
    xD_q(x)        &= 0.5\cdot xD_{\rm sea}(x) +xD_{2u}(x),\\
    xD_{\bar q}(x) &= 0.5\cdot xD_{\rm sea}(x) +xD_{d}(x),\\
    xD_{\rm sea}(x)&= 0.6733449216\cdot x^{-0.2}(1-x)^{7.0},\\
    xD_{2u}(x)     &= 2.1875000000\cdot x^{ 0.5}(1-x)^{3.0},\\    
    xD_{d}(x)      &= 1.2304687500\cdot x^{ 0.5}(1-x)^{4.0},
  \end{split}
\end{equation}

\begin{figure}[!ht]
  \centering
  {\epsfig{file=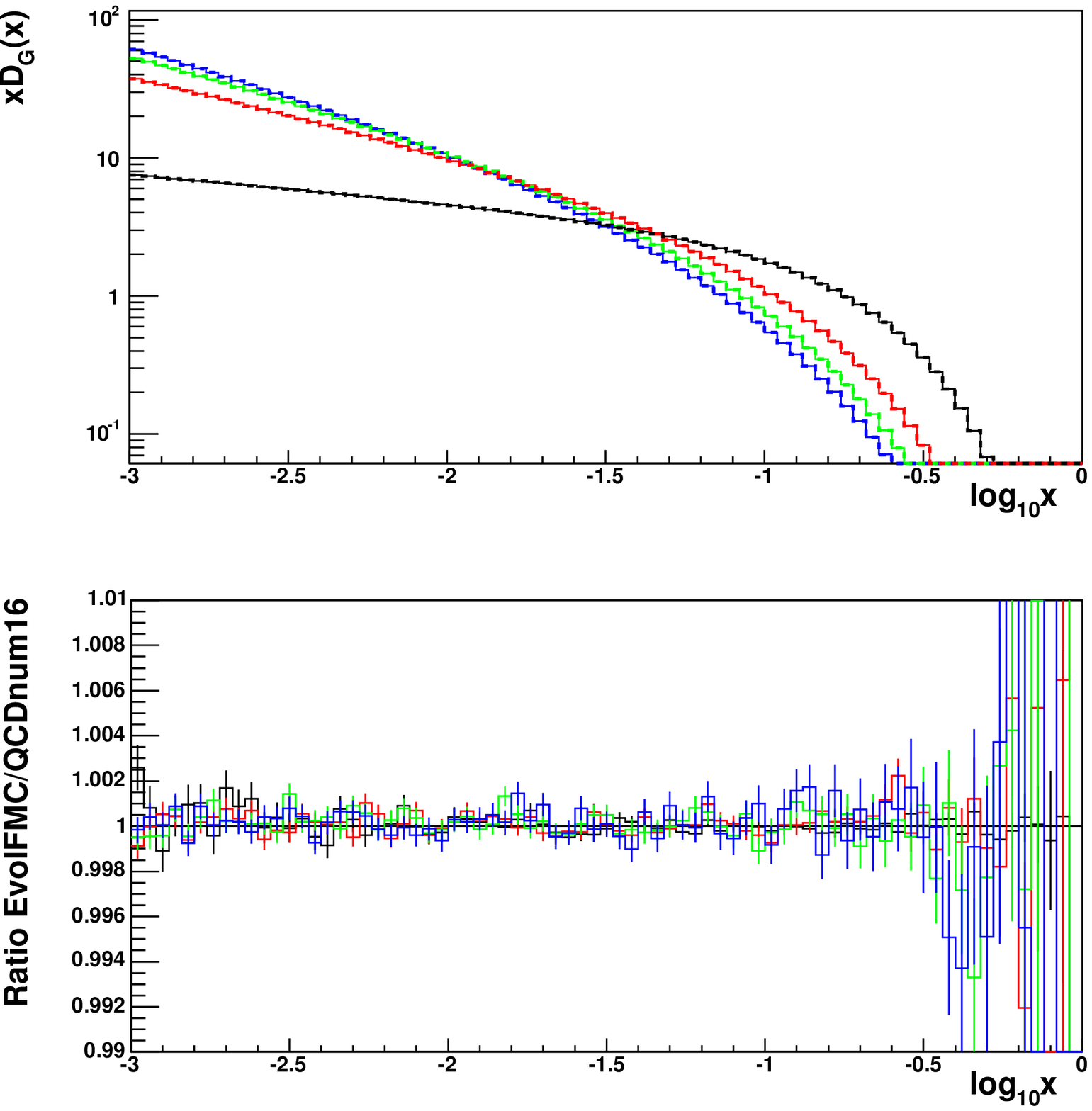,width=100mm,height=100mm}}
  \caption{\sf
    The upper plot shows the gluon distribution $xD_G(x,Q_i)$ 
    evolved from $Q_0=1\,$GeV (black) to $Q_i=10$ (red), 
    $100$ (green) and $1000$ (blue) GeV,
    obtained in the LO approximation from {\tt EvolFMC} (solid lines) and  
    {\tt QCDnum16} (dashed lines), 
    while the lower plot shows their ratio.
    }
  \label{fig:Gs_LO}
\end{figure}
%
\begin{figure}[!ht]
  \centering
  {\epsfig{file=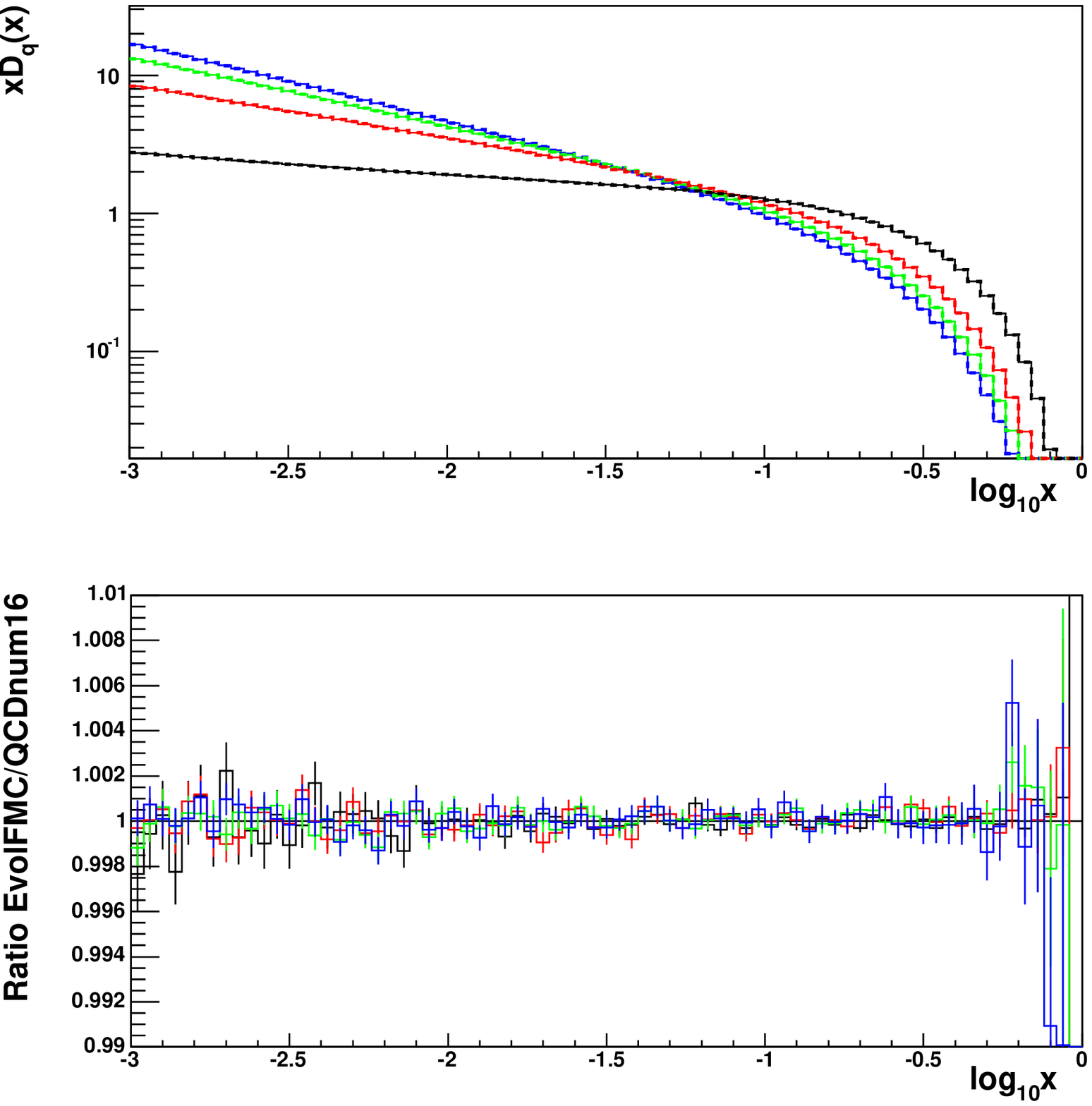,width=100mm,height=100mm}}
  \caption{\sf
    The upper plot shows the singlet quark distribution $D_{\bar{q}})$ 
    evolved from $Q_0=1\,$GeV (black) to $Q_i=10$ (red), 
    $100$ (green) and $1000$ (blue) GeV,
    obtained in the LO approximation from {\tt EvolFMC} (solid lines) and  
    {\tt QCDnum16} (dashed lines), 
    while the lower plot shows their ratio.
    }
  \label{fig:Qs_LO}
\end{figure}

The first results of these comparisons for the LO evolution were presented in 
Ref.~\cite{Jadach:2003bu}. They showed a $0.2\%$ discrepancy for the
gluon distributions between {\tt EvolFMC} and {\tt QCDnum16}.
This numerical bias is eliminated in this paper.
In Figs.~\ref{fig:Gs_LO}--\ref{fig:Qs_LO}
we show the resulting gluon and quark distributions evolved
to $Q=10$, $100$, $1000\,$ GeV in the LO approximation.
As one can see, these two calculations agree to within $0.1\%$ for the gluon 
as well as for the quark-singlet distributions.
The origin of the previous $0.2\%$ discrepancy for gluon was identified
as a result of too high values of 
the dummy IR cut-offs: $\epsilon = \bar{\epsilon} = 10^{-3}$.
The new results have been obtained for $\epsilon = \bar{\epsilon} = 10^{-4}$.
In the small-$x$ region ($x < 0.1$), we have found a similar agreement with 
the program {\tt  APCheb33}~\cite{APCheb33}, 
which uses the Chebyshev-polynomial technique to solve the DGLAP equations.

\begin{figure}[!ht]
  \centering
  {\epsfig{file=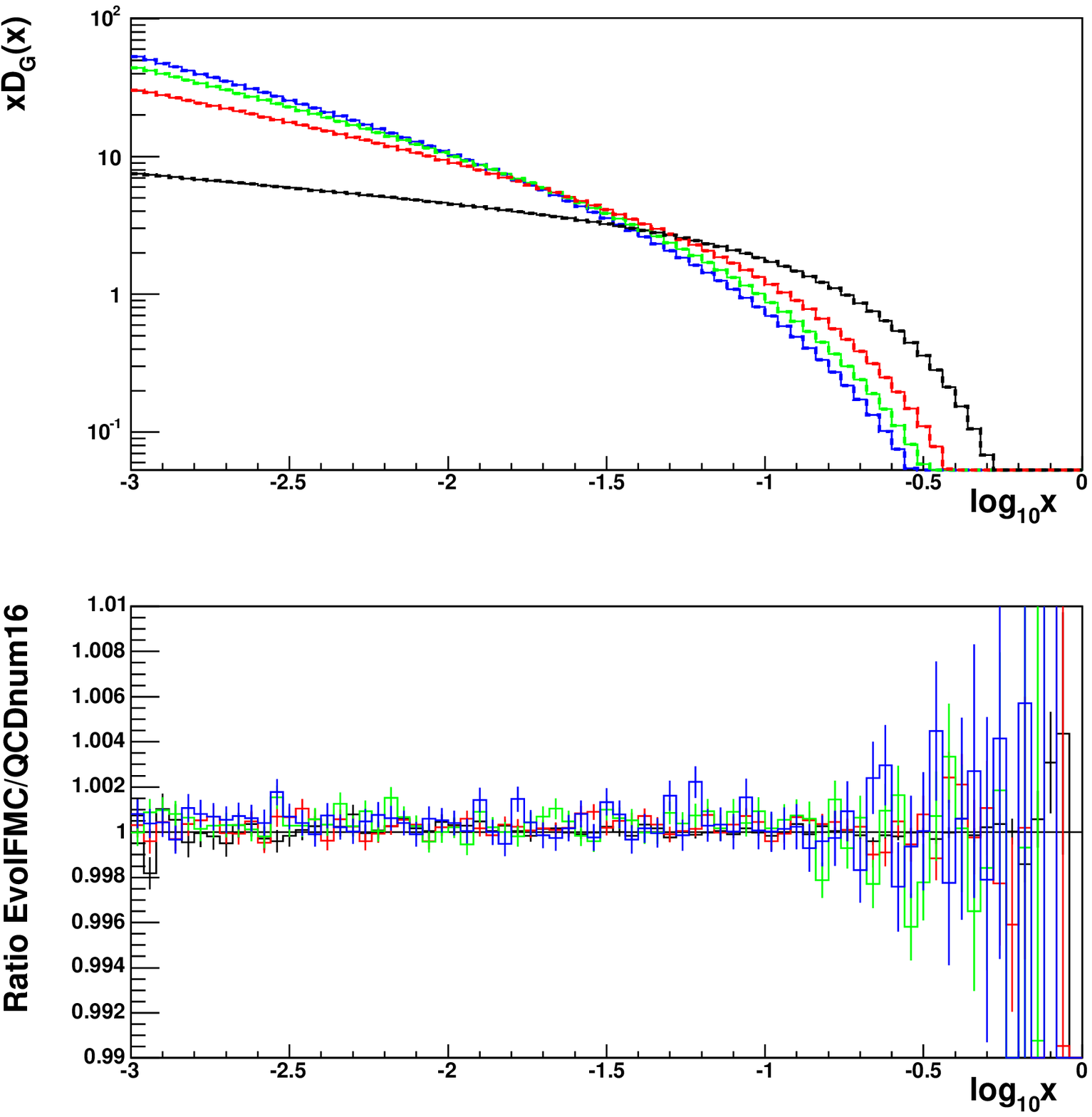,width=100mm,height=100mm}}
  \caption{\sf
    The upper plot shows the gluon distribution $xD_G(x,Q_i)$ 
    evolved from $Q_0=1\,$GeV (black) to $Q_i=10$ (red), 
    $100$ (green) and $1000$ (blue) GeV,
    obtained in the NLO approximation from {\tt EvolFMC} (solid lines) and  
    {\tt QCDnum16} (dashed lines), 
    while the lower plot shows their ratio.
    }
  \label{fig:Gs_NLO}
\end{figure}
%
\begin{figure}[!ht]
  \centering
  {\epsfig{file=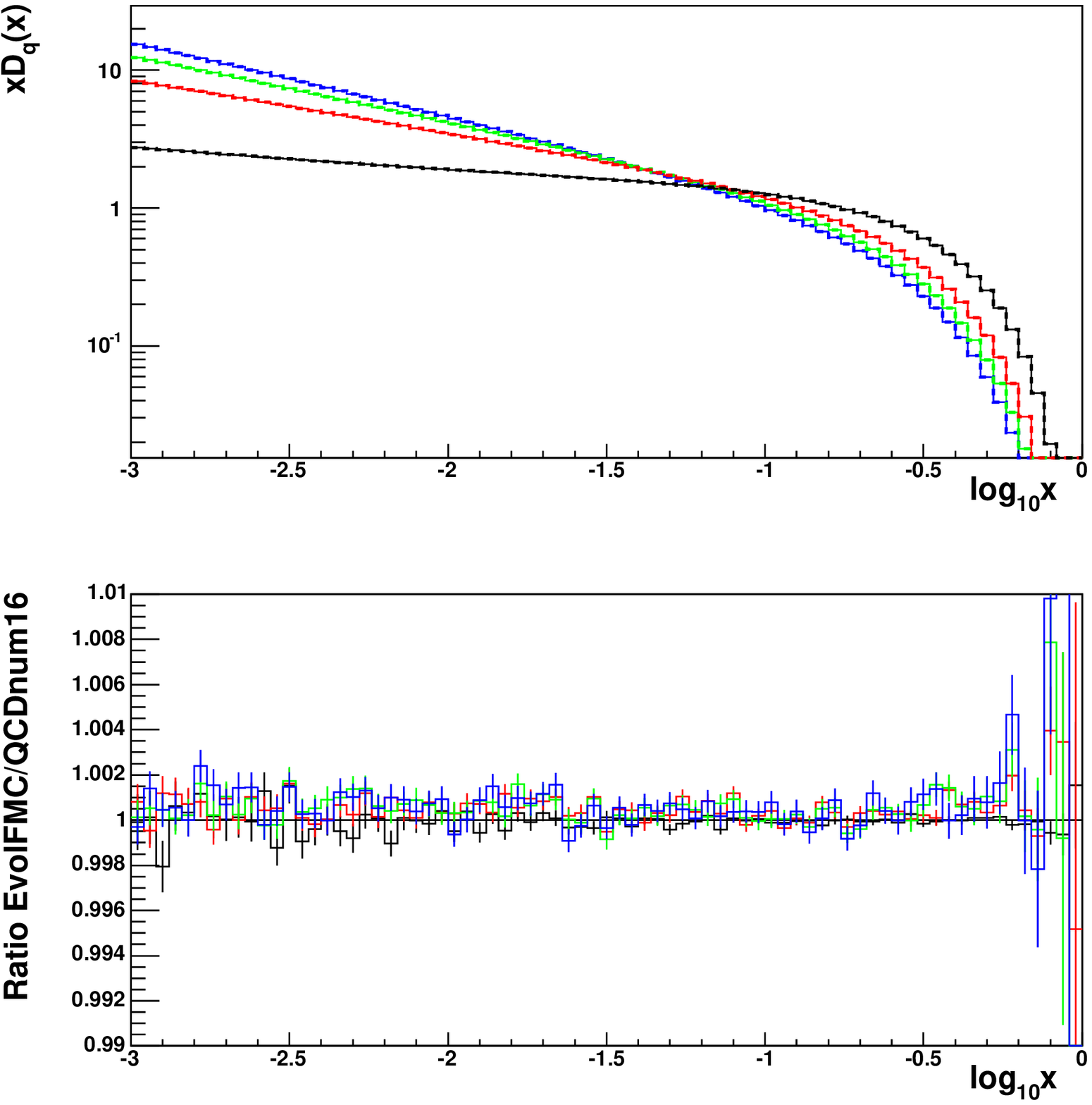,width=100mm,height=100mm}}
  \caption{\sf
    The upper plot shows the singlet quark distribution $xD_{\bar{q}}$ 
    evolved from $Q_0=1\,$GeV (black) to $Q_i=10$ (red), 
    $100$ (green) and $1000$ (blue) GeV,
    obtained in the NLO approximation from {\tt EvolFMC} (solid lines) and  
    {\tt QCDnum16} (dashed lines), 
    while the lower plot shows their ratio.
    }
  \label{fig:Qs_NLO}
\end{figure}

In Figs.~\ref{fig:Gs_NLO}--\ref{fig:Qs_NLO} we present the results of similar
comparisons for the NLO evolution. In the NLO, there is some ambiguity
in calculation of the running $\alpha_s$. In {\tt EvolFMC} we use a
definition of $\alpha_s$ given in Appendix~\ref{sec:kernelsNLO}. 
However, in the original version of {\tt QCDnum16} a
different definition of $\alpha_s$ at the NLO was employed. For the sake
of our comparisons, we have replaced in the {\tt QCDnum16} code the routine
for $\alpha_s$ evaluation with the appropriate routine from {\tt EvolFMC}.
We have checked for several values of $Q^2$, that after replacement 
the two programs give numerically  the same values of $\alpha_s(Q^2)$. 
The NLO results for the gluon and quark-singlet distributions from 
{\tt EvolFMC} and {\tt QCDnum16} agree within $\sim 0.1\%$, as in the LO case.
Again, for $x < 0.1$ we have found a similar agreement with {\tt APCheb33}.

The MC calculation for the NLO evolution is, of course, slower and less 
efficient than for the LO one but not very much. 
In comparison with the LO results given above, the NLO results 
were obtained for the statistics about 2 times higher and their computation 
required about 3 times more CPU time. 

All the above results were obtained in the weighted-event mode of 
{\tt EvolFMC}.  In the LO case the weighted events can be turned into the
unweighted (weight $=1$) ones without difficulty -- the event weight
is well-behaved, non-negative and bounded from above.  At the NLO,
however, the situation is problematic. As was described in
Section~\ref{sec:z1}, the $P_{FG}$ and $P_{GF}$ splitting functions at
the NLO acquire logarithmic singularities at $z=1$.
In our MC algorithm this leads to large positive weights for the
$F\rightarrow G$ transitions and to negative weights for the
$G\rightarrow F$ transitions in the region of $z\gtrsim 0.95$. While
the problem of large positive weights can be cured with appropriate
importance sampling, there is no technical method to turn
negative-weight events into unweighted events.  The current version of
{\tt EvolFMC} implements only the weighted-event solution for the NLO
DGLAP evolution.  These problems will be addressed in our future
works.

\section{Summary and outlook}
\label{sec:summary}

In this paper we have presented in detail two Markovian MC algorithms
for solving the DGLAP equations up to the NLO accuracy. 
The one of them is based directly on the evolution of the PDFs 
and the other one instead of pure PDFs uses the parton-momentum distributions. 
The latter algorithm is more efficient due to the momentum-conservation sum 
rules. The evolution is done in parton flavour space, in the current
version for gluons and three light quarks but it can be easily extended 
to include more flavours.
Both the above algorithms have been implemented in the MC generator 
{\tt EvolFMC} (written in C$++$). This program has been cross-checked
against two independent non-MC programs: {\tt QCDnum16} and {\tt APCheb33}.
The numerical tests show that with today computer CPU power the Markovian MC 
is able to solve efficiently and precisely (to the per-mil level) 
the QCD evolution equation up to the NLO.
Therefore, it can be used to cross-check other, non-MC methods or
even as an alternative to them. As was pointed out in Introduction,
the MC method is not competitive with other techniques in terms
of the CPU time but is has certain advantages that may be important
in some cases -- it is usually less biased and more stable numerically
as well as more flexible for possible extensions.  

So far we have included only light (massless) quarks in our
MC algorithm, however, extending it to heavy quarks does not pose
any problem. It can be realized either by simple rejection
of extra massless quarks below mass thresholds or by some importance
sampling that accounts for these thresholds.
Also extension to the NNLO seems straightforward -- one only needs
to implement the NNLO evolution kernels~\cite{Moch:2004pa,Vogt:2004mw}.   
A more serious problem concerns the divergences of the NLO kernels 
at $z\rightarrow 1$ which for some transitions lead to negative weights. 
This indicates that some resummation in this region might be necessary.  

As was mentioned in Introduction, we do not consider the Markovian
MC algorithm described in this paper to be only a tool for solving 
numerically the evolution equations for parton densities. 
It can be also used as the basis for constructing the FSR parton shower 
MC event generator which would generate physical events in terms of particle 
flavours and four-momenta. 
This algorithm cannot be directly used for the ISR evolution (parton
shower), since in that case one needs to impose some energy-momentum
constraints on the partons entering the hard process. However, it
can be used as a starting point for developing any kind of constraint
MC algorithms as well as it can play a role of a testing tool
for corresponding MC programs, 
see Refs.~\cite{raport04-06,raport04-07,zinnowitz04,herwig-cmc}.

Yet another way of development of the above Markovian MC algorithm can
go into the direction of solving the CCFM equations~\cite{CCFM}.
In this case one deals with the so-called unintegrated parton
distribution functions, which in addition to $x$ and $Q^2$ depend
on the partonic transverse momenta $k_T$. This may be important for
better modeling physical events in deep inelastic scattering
as well as hadron--hadron collisions, see e.g. Ref.~\cite{Andersson:2002cf}.
We have already implemented the so-called one-loop approximation
of the CCFM equation~\cite{Marchesini:1991zy,Jung:2000hk,Kwiecinski:2002} which
will be reported in our forthcoming paper~\cite{raport05-03}.

\appendix

\section{QCD kernels at NLO}
\label{sec:kernelsNLO}

A general parton--parton transition matrix for a gluon and three quark 
flavours $(d,\,u,\,s)$ can be written as
\begin{equation}
\Pbf(\alpha_s,z) =  
\left[
\begin{array}{c|ccc|ccc}
  P_{G\from G}, &
  P_{G\from d}, &  P_{G\from u}, &  P_{G\from s}, &
  P_{G\from\bar{d}}, & P_{G\from\bar{u}}, & P_{G\from\bar{s}} 
\\ \hline
  P_{d\from G}, &
  P_{d\from d}, & P_{d\from u}, &  P_{d\from s}, &
  P_{d\from\bar{d}}, & P_{d\from\bar{u}}, & P_{d\from\bar{s}}
\\
  P_{u\from G}, &
  P_{u\from d}, & P_{u\from u}, &  P_{u\from s}, &
  P_{u\from\bar{d}}, & P_{u\from\bar{u}}, & P_{u\from\bar{s}}
\\ 
  P_{s\from G}, &
  P_{s\from d}, & P_{s\from u}, &  P_{s\from s}, &
  P_{s\from\bar{d}}, & P_{s\from\bar{u}}, & P_{s\from\bar{s}}
\\ \hline
  P_{\bar{d}\from G}, &
  P_{\bar{d}\from d}, & P_{\bar{d}\from u}, &  P_{\bar{d}\from s}, &
  P_{\bar{d}\from\bar{d}}, & P_{\bar{d}\from\bar{u}}, & P_{\bar{d}\from\bar{s}}
\\
  P_{\bar{u}\from G}, &
  P_{\bar{u}\from d}, & P_{\bar{u}\from u}, &  P_{\bar{u}\from s}, &
  P_{\bar{u}\from\bar{d}}, & P_{\bar{u}\from\bar{u}}, & P_{\bar{u}\from\bar{s}}
\\
  P_{\bar{s}\from G}, &
  P_{\bar{s}\from d}, & P_{\bar{s}\from u}, &  P_{\bar{s}\from s}, &
  P_{\bar{s}\from\bar{d}}, & P_{\bar{s}\from\bar{u}}, & P_{\bar{s}\from\bar{s}}
\\
\end{array} 
\right],
\end{equation}
where $P_{J\from I} \equiv P_{J\from I}(\alpha_s,z)$. 
At the NLO, the kernels can be decomposed as follows
\begin{equation}
\Pbf(\alpha_s,z) = \frac{\alpha_s(t)}{2\pi}\, \Pbf^{(0)}(z) + 
                   \left(\frac{\alpha_s(t)}{2\pi}\right)^2\, \Pbf^{(1)}(z),
\end{equation}
where the NLO QCD coupling in the $\overline{MS}$-scheme is 
\begin{equation}
\begin{split}
 \alpha_s(t) & = \alpha^{(0)}_s(t)\,
 \left\{ 1 - \alpha^{(0)}_s(t)\, 
 \frac{b_1}{b_0}\, \ln\left(2[t - \ln\Lambda_{\overline{MS}}]\right)\right\}, 
\\
b_0 = \frac{\beta_0}{4\pi},& \hspace{4mm} 
b_1 = \frac{\beta_1}{(4\pi)^2}, \hspace{4mm}
\beta_0 = 11 - \frac{2}{3}\,n_f\,, \hspace{4mm}
\beta_1 = 102 - \frac{38}{3}\,n_f\,, 
\end{split}
\end{equation}
and $t=\ln Q$ ($n_f$ is the number of active flavours).

The LO kernel matrix takes a simple form
\begin{equation}
\Pbf^{(0)}(z)=  
\left[
\begin{array}{c|ccc|ccc}
  P_{GG}^{(0)}, &
  P_{GF}^{(0)}, & P_{GF}^{(0)}, & P_{GF}^{(0)}, & 
  P_{GF}^{(0)}, & P_{GF}^{(0)}, & P_{GF}^{(0)} 
\\ \hline
  P_{FG}^{(0)}, &
  P_{FF}^{(0)}, & 0 , &  0, & 0 , & 0, & 0
\\ 
  P_{FG}^{(0)}, &
   0, & P_{FF}^{(0)}, & 0 , &  0, & 0 , & 0
\\ 
  P_{FG}^{(0)}, &
  0 , &  0, & P_{FF}^{(0)}, & 0 , & 0, & 0
\\ \hline 
  P_{FG}^{(0)}, &
  0 , &  0, & 0 , & P_{FF}^{(0)}, & 0, & 0
\\ 
  P_{FG}^{(0)}, &
  0 , &  0, & 0 , & 0, & P_{FF}^{(0)}, & 0
\\
  P_{FG}^{(0)}, &
  0 , &  0, & 0 , & 0, & 0, & P_{FF}^{(0)}
\\
\end{array} 
\right],
\end{equation}
where 
\begin{equation}
\begin{split}
&  P^{(0)}_{GG}(z)
   =2C_A\, \Big[ \frac{1}{(1-z)_+} -2 +z(1-z) +\frac{1}{z}\Big]
                       +\frac{11C_A -4T_f}{6}\,\delta(1-z),
\\&
   P^{(0)}_{FG}(z)=
   T_R\, [z^2+(1-z)^2],
\\&
   P^{(0)}_{GF}(z)=
   C_F\,  \frac{1+(1-z)^2}{z},
\\&
   P^{(0)}_{FF}(z)=
   C_F\,  \Big[ \frac{1+z^2}{(1-z)_+} +\frac{3}{2}\,\delta(1-z) \Big]\,,
\end{split}
\end{equation}
and the colour-group factors are: $C_A=N_c=3,\; C_F=(N_c^2-1)/2N_c=4/3,\; T_R=1/2$.

The NLO contribution to the kernel matrix can be expressed in the following 
form
\begin{equation}
\Pbf^{(1)}(z)=  
\left[
\begin{array}{c|ccc|ccc}
  P_{GG}^{(1)}, &
  P_{GF}^{(1)}, & P_{GF}^{(1)}, & P_{GF}^{(1)}, & 
  P_{GF}^{(1)}, & P_{GF}^{(1)}, & P_{GF}^{(1)} 
\\ \hline
  P_{FG}^{(1)}, &
  P_{qq}^{V+S\,(1)}, &   P_{qq}^{S\,(1)}, &   P_{qq}^{S\,(1)}, & 
  P_{q\bar{q}}^{V+S\,(1)}, & P_{q\bar{q}}^{S\,(1)}, & P_{q\bar{q}}^{S\,(1)}
\\ 
  P_{FG}^{(1)}, &
  P_{qq}^{S\,(1)}, &   P_{qq}^{V+S\,(1)}, &   P_{qq}^{S\,(1)}, & 
  P_{q\bar{q}}^{S\,(1)}, & P_{q\bar{q}}^{V+S\,(1)}, & P_{q\bar{q}}^{S\,(1)}
\\ 
  P_{FG}^{(1)}, &
  P_{qq}^{S\,(1)}, &   P_{qq}^{S\,(1)}, &   P_{qq}^{V+S\,(1)}, & 
  P_{q\bar{q}}^{S\,(1)}, & P_{q\bar{q}}^{S\,(1)}, & P_{q\bar{q}}^{V+S\,(1)}
\\ \hline 
  P_{FG}^{(1)}, &
  P_{q\bar{q}}^{V+S\,(1)}, & P_{q\bar{q}}^{S\,(1)}, & P_{q\bar{q}}^{S\,(1)}, &
  P_{qq}^{V+S\,(1)}, &   P_{qq}^{S\,(1)}, &   P_{qq}^{S\,(1)} 
\\ 
  P_{FG}^{(1)}, &
  P_{q\bar{q}}^{S\,(1)}, & P_{q\bar{q}}^{V+S\,(1)}, & P_{q\bar{q}}^{S\,(1)}, & 
  P_{qq}^{S\,(1)}, &   P_{qq}^{V+S\,(1)}, &   P_{qq}^{S\,(1)}
\\
  P_{FG}^{(1)}, &
  P_{q\bar{q}}^{S\,(1)}, & P_{q\bar{q}}^{S\,(1)}, & P_{q\bar{q}}^{V+S\,(1)}, &
  P_{qq}^{S\,(1)}, &   P_{qq}^{S\,(1)}, &   P_{qq}^{V+S\,(1)} 
\\
\end{array} 
\right],
\end{equation}
where $P_{IJ}^{(1)}\equiv P_{IJ}^{(1)}(z)$ and we have used a short-hand notation
$P_{IJ}^{V+S\,(1)} \equiv P_{IJ}^{V\,(1)} + P_{IJ}^{S\,(1)}$. The above matrix
can be simplified by exploiting the identity (\ref{eq:dglap7b})
\begin{equation}
P_{qq}^{S\,(1)} = P_{q\bar{q}}^{S\,(1)},
\end{equation}
which is true up to the NLO. However, we prefer to keep a more general form of 
the kernel matrix which can be useful for some tests and possible future
extensions. The above kernel matrices can be easily extended to include more
quark flavours.

The non-singlet and singlet-quark kernels can be expressed in terms 
of the basic NLO splitting functions $P_+,\,P_-$ and $P_{FF}$, 
defined in Refs.~\cite{Furmanski:1980cm,Curci:1980uw}, as follows:
\begin{equation}
P_{qq}^{V\,(1)} = \frac{1}{2} \left[P_+^{(1)} + P_-^{(1)}\right], \hspace{2mm}
P_{q\bar{q}}^{V\,(1)} = \frac{1}{2} \left[P_+^{(1)} - P_-^{(1)}\right], \hspace{2mm}
P_{qq}^{S\,(1)} = \frac{1}{2n_f} \left[P_{FF}^{(1)} - P_+^{(1)}\right].
\end{equation}

Finally, all the elements of the above kernel matrix can be calculated (e.g.\ numerically)
from the six basic NLO splitting functions of Refs.~\cite{Furmanski:1980cm,Curci:1980uw}
\begin{equation}
\left[ P_+^{(1)},\,P_-^{(1)},\,P_{FF}^{(1)},\,P_{FG}^{(1)},\,P_{GF}^{(1)},\,P_{GG}^{(1)}
\right].
\end{equation}
These basic splitting functions are given at the NLO by the following expressions:
\begin{equation}
P^{(1)}_{\pm}(z,\epsilon) =
\frac{A^{(1)}_\pm(z)}{1-z}\,\Theta (1-z-\epsilon)\,+\,B^{(1)}_\pm(z)\,+\,
\left[C^{(1)}_S+A^{(1)}_S\ln\epsilon\right]\,\delta(1-z),
\end{equation}
\begin{equation}
P^{(1)}_{FF}(z,\epsilon)  =
\frac{A^{(1)}_S}{1-z}\,\Theta (1-z-\epsilon)\,+\,B^{(1)}_{S}(z)\,+\,
\left[C^{(1)}_S+A^{(1)}_S\ln\epsilon\right]\,\delta(1-z), 
\end{equation}
\begin{equation}
P^{(1)}_{GG}(z,\epsilon) =
\frac{A^{(1)}_G}{1-z}\,\Theta (1-z-\epsilon)\,+\,B^{(1)}_{G}(z)\,+\,
\left[C^{(1)}_G+A^{(1)}_G\ln\epsilon\right]\,\delta(1-z),
\end{equation}
\begin{equation}
\begin{split}
P^{(1)}_{FG}(z) &=
\frac{1}{2}\, C_F\,T_R\bigg\{
4-9z+(4z-1)\ln z +(2z-1)\ln^2 z+4\ln(1-z) 
\\&~~~~~~~~~~~~+\bigg[
10-\frac{2}{3}\pi^2
+2\ln^2\left(\frac{1- z}{z}\right)-4\ln\left(\frac{1- z}{z}\right)
\bigg] \left[z^2+(1-z)^2\right]
\bigg\} \\
& + \frac{1}{2}\, C_A\,T_R\,\bigg\{
\frac{182}{9}+\frac{14}{9}z+\frac{40}{9z}+\left(\frac{136}{3}z-\frac{38}{3}\right)\ln z
-4\ln(1-z)
\\&~~~~~~~~~~~~~
-(2+8z)\ln^2 z 
+\,2\left[z^2+(1+z)^2 \right]\, S_2(z)\,
+\,
\bigg[
\frac{\pi^2}{3}-\frac{218}{9}
\\&~~~~~~~~~~~~~
+\frac{44}{3}\ln z-\ln^2 z 
+\,4\ln(1-z)-2\ln^2(1-z)
\bigg]\left[z^2+(1-z)^2 \right]
\bigg\},
\end{split}
\end{equation}
\begin{equation}
\begin{split}
P^{(1)}_{GF}(z) & =
C_F^2\,\bigg\{
-\frac{5}{2}  - \frac{7}{2}z - 2z\ln(1 - z) -
  \frac{1 + ( 1 - z)^2}{z} \,\ln(1 - z)
     \left[3+ {\ln(1 - z)} \right] 
\\&~~~~~~
+\left( 2 + \frac{7}{2}z \right)\ln z -
  \left(1 - \frac{1}{2}z \right)\ln^2 z
\bigg\} 
\\& + C_F\, C_A\,\bigg\{
\frac{28}{9} + \frac{65}{18}z + \frac{44}{9}z^2 -
  \left(12 + 5z + \frac{8}{3}z^2 \right)\ln z+\left( 4 + z \right)\ln^2 z 
\\&~~~~~~~~~~
+\, 2 z \ln(1 - z)\,+\,
  \frac{1 + (1 - z)^2}{z}\,
     \bigg[ \frac{1}{2} - \frac{{\pi }^2}{6} +
     \frac{11}{3}\ln(1 - z) 
\\&~~~~~~~~~~
+\,\ln^2(1 - z)- 2\ln z\, \ln(1 - z) +\frac{1}{2}\ln^2 z\,
       \bigg]\, -\, \frac{1 + ( 1 + z)^2}{z}\,S_2(z) \bigg\} 
\\& 
- C_F\, T_f\,\bigg\{
\frac{4}{3}z\,+\, \frac{1 + ( 1 - z)^2}{z}
     \left[ \frac{20}{9} + \frac{4}{3}\ln(1-z) \right]
\bigg\},
\end{split}
\end{equation}
where 
\begin{equation}
\begin{split}
C^{(1)}_S &= 
C_F^2\,\left(\frac{3}{8}-\frac{\pi^2}{2}+6\zeta_3\right)+
\frac{1}{2}C_F\,C_A\,\left(\frac{17}{12}+\frac{11\pi^2}{9}-6\zeta_3\right)
-C_F\,T_f\,\left(\,\frac{1}{6}+\frac{2\pi^2}{9}\right),
\\
A^{(1)}_S &= 
C_F\left\{C_A\,\left(\frac{67}{9}-\frac{\pi^2}{3}\right)-\frac{20}{9}T_f\right\},
\\
C^{(1)}_G &=
C_A^2\left(\frac{8}{3}+3\zeta_3\right) -\left(\frac{4}{3}C_A+C_F\right)T_f,
\\
A^{(1)}_G &= C_A\left\{
C_A\!\left(\frac{67}{9}-\frac{\pi^2}{3}\right)-\frac{20}{9}T_f
\right\},
\end{split}
\end{equation}
with $\zeta_3\equiv\zeta(3)\approx 1.2020569$.

The non-singlet coefficients take the form 
\begin{equation}
\begin{split}
A^{(1)}_\pm(z) & =
C_F^2\left[\!-2(1+z^2)\ln z\ln(1-z)-{3}\ln z\right]
\\&
+ \frac{1}{2}\, C_F\, C_A \left[\ln^2 z +\frac{11}{3}\ln z
+\frac{67}{9}-\frac{\pi^2}{3}\right](1+z^2)
\\&
- C_F\, T_f\, \frac{2}{3}\left[\ln z+\frac{5}{3}\right](1+z^2),
\end{split}
\end{equation}
\begin{equation}
\begin{split}
B^{(1)}_\pm(z) & = 
C_F^2\left[- 2z\ln z-\frac{1}{2}(1+z)\ln^2 z - 5(1-z)\,\pm\, P_A(z)\right]
\\&
+ \frac{1}{2}\, C_F\, C_A\left[2(1+z)\ln z+\frac{40}{3}(1-z)\,\mp\, P_A(z)\right]
- C_F\, T_f\,\frac{4}{3} (1-z),
\end{split}
\end{equation}
where
\begin{equation}
P_A(z) = 2\,\frac{1+z^2}{1+z}\, S_2(z)\,+\,2(1+z)\ln z\,+\,4(1-z)\,.
\end{equation}
Notice that, as it should be, $A^{(1)}_\pm(1)=A^{(1)}_S$. 
The function $S_2$ is defined in the interval $(0,1]$
\begin{equation}
S_2(z) = \int\limits_{z/(1+z)}^{1/(1+z)}\frac{dy}{y}\ln\frac{1-y}{y}
       = -2\,\mbox{\rm Li}_2(-z)\,+\,\frac{1}{2}\ln^2 z\,-\,2\ln z \ln(1+z) 
         \,-\,\frac{\pi^2}{6},
\end{equation}
where the dilogarithm function ${\rm Li}_2$ is given by
\begin{equation}
\mbox{\rm Li}_2(z)\,=\,\sum_{k=1}^{\infty}\frac{z^k}{k^2}\,=\,
\int\limits_z^0\frac{dt}{t}\ln(1-t),
\end{equation}
with a branch point discontinuity on a complex plane along $[1,+\infty]$.

For the singlet and gluon coefficients we have
\begin{equation}
\begin{split}
B_S^{(1)}(z) &=
C_F^2\,\bigg\{
-1 + z + \frac{1}{2} \left[ 1 - 3z \,-\,(1 + z)\,\ln z \right] \ln z 
\\&~~~~~~~~~~~
-2 \left( \frac{1+z^2}{1 - z} \right) \left[\frac{3}{4}  + \ln (1 - z) \right] \ln z
+ \,2\,\frac{1 + z^2}{1 + z}\,S_2(z)
\bigg\}
\\&
+ C_F\, C_A\,\bigg\{
-\left( \frac{67}{18} - \frac{\pi^2}{6} \right)(1 + z)\,+\, \frac{14}{3}(1-z)
\\&~~~~~~~~~~~~
+\frac{1}{2} \left( \frac{1+z^2}{1 - z} \right) \left( \frac{11}{3} + \ln z \right) \ln z
-\,\frac{1 + z^2}{1 + z}\,S_2(z)
\bigg\}
\\&
+ C_F\, T_f\,\bigg\{
-\frac{16}{3} + \frac{10}{9}(1 + z) +\frac{40}{9z}+ \frac{40z}{3} - \frac{112 z^2}{9} -
  \frac{2}{3} \left( \frac{1 + z^2}{1 - z} \right)\ln z
\\&~~~~~~~~~~~~  
+ \left( 2 + 10\,z + \frac{16}{3} z^2 \right) \ln z - 2\,( 1 + z ) \,\ln^2 z 
\bigg\},
\end{split}
\end{equation}
\begin{equation}
\begin{split}
B_G^{(1)}(z) &=
C_A^2\bigg\{ \frac{27}{2}(1-z)+\frac{67}{9}(z^2-z^{-1})
- \frac{1}{3}\left(25 - 11 z+ 44 z^2 \right)\ln z
\\&~~~~~
+\,4(1+z)\ln^2 z\,+\,\left[\ln^2 z-4\ln z\ln(1- z)\right]\,\frac{1}{1-z}
\\&~~~~~
+\,\left[\ln^2 z-4\ln z\ln(1- z)+\frac{67}{9}-\frac{\pi^2}{3}\right]
\left[\frac{1}{z} -2 +z(1-z)\right]
\\&~~~~~
+\,2\left(\frac{1}{1+z}-\frac{1}{z} -2 -z - z^2\right)\,S_2(z)
\bigg\}
\\&
+ C_A\, T_f\,\bigg\{
2(1-z)+\frac{26}{9}(z^2-z^{-1})-\frac{4}{3}(1+z)\ln z-\frac{20}{9}
\left[\frac{1}{z}-2+z(1-z)\right]
\bigg\}
\\&
+C_F\, T_f\,\bigg\{
8(z-2)+\frac{20}{3} z^2 +\frac{4}{3z} -(6+10z) \ln z -2(1+z) \ln^2 z
\bigg\}.
\end{split}
\end{equation}
It is easy to check that the above functions have no singularity at $z=1$.

It is worth noticing that the non-singlet quark kernels have particularly 
simple analytical representations
\begin{equation}
\begin{split}
P^{V\,(1)}_{qq}(z,\epsilon) & =
\left[C^{(1)}_S+A^{(1)}_S\ln\epsilon\right]\,\delta(1-z) \,+\,
\frac{A^{(1)}_+(z)}{1-z}\,\Theta (1-z-\epsilon)
\\&
 +\, C_F^2\left[- 2z\ln z-\frac{1}{2}(1+z)\ln^2 z - 5(1-z)\right]
\\&
+ \frac{1}{2}\, C_F\, C_A\left[2(1+z)\ln z+\frac{40}{3}(1-z)\right]
- C_F\, T_f\,\frac{4}{3} (1-z),
\end{split}
\end{equation}
\begin{equation}
P^{V\,(1)}_{q\bar{q}}(z) = C_F\left(C_F - \frac{1}{2}\,C_A \right)P_A(z). 
\end{equation}
Using these analytical formulae in numerical evaluations
of the kernels $P^{V\,(1)}_{qq}$ and $P^{V\,(1)}_{q\bar{q}}$ 
can be faster and more stable numerically than computing them indirectly from 
the splitting functions $P_{\pm}^{(1)}$.

In the Monte Carlo implementation, the above kernel matrices are used for
generation of real-parton radiation, i.e.\ for $z < 1 - \epsilon$. 
This has to be compensated by the appropriate Sudakov form factor
summing up virtual and soft-parton corrections, i.e.\ terms proportional
to $\delta(1-z)$. At the NLO, the Sudakov exponent $\Phi_K$ takes the form
\begin{equation}
\Phi_K(t_2,t_1) = 2 \int\limits_{t_1}^{t_2} dt \, 
\left\{
   \frac{\alpha_s(t)}{2\pi}\, P_{KK}^{\delta\,(0)}(z) + 
   \left(\frac{\alpha_s(t)}{2\pi}\right)^2\, P_{KK}^{\delta\,(1)}(z)
\right\},
\end{equation}
where the NLO $\alpha_s(t)$ has to be taken for both terms. The factor
of $2$ in front of the integral is due to the fact that our evolution ``time''
is $t=\ln Q$. Integrating over 
$\tau = \ln(t - \ln\Lambda)$, we obtain for gluons
\begin{equation}
\begin{split}
\Phi_G(\tau_2,\tau_1) = \frac{2}{\beta_0} \bigg\{ &
\left[ \chi_0(\tau_2) - \chi_0(\tau_1) \right]\,
\left[ 2C_A \ln\frac{1}{\epsilon} - \frac{11}{6}C_A + \frac{2}{3}T_f \right]
\\ + &
\left[\chi_1(\tau_2) - \chi_1(\tau_1) \right]\,
\left[A_G^{(1)}\ln\frac{1}{\epsilon} - C_G^{(1)} \right] \bigg\}
\end{split}
\end{equation}
and for fermions (both quarks and anti-quarks)
\begin{equation}
\begin{split}
\Phi_F(\tau_2,\tau_1) = \frac{2}{\beta_0} \bigg\{ &
\left[ \chi_0(\tau_2) - \chi_0(\tau_1) \right]\,
\left[ 2C_F \ln\frac{1}{\epsilon} - \frac{3}{2}C_F \right]
\\ + & 
\left[\chi_1(\tau_2) - \chi_1(\tau_1) \right]\,
\left[A_S^{(1)}\ln\frac{1}{\epsilon} - C_S^{(1)}\right] \bigg\},
\end{split}
\end{equation}
where
\begin{equation}
\chi_0(\tau)= \tau + \frac{\beta_1}{2\beta_0^2}(\tau + \ln 2 + 1)\,e^{-\tau}\,,
\end{equation}
and
\begin{equation}
\begin{split}
\chi_1(\tau) =  - \frac{1}{\beta_0} \Bigg\{ 
\frac{\beta_1}{2\beta_0^2}\bigg( &
\frac{\beta_1}{6\beta_0^2}
\left[\tau^2 + 2\left(\ln 2 + \frac{1}{3}\right)\left(\tau + \frac{1}{3}\right)
 + \ln^2 2 \right] \,e^{-\tau}
\\ & 
 - \tau - \ln 2 - \frac{1}{2}\bigg)\,e^{-\tau}  + 1\Bigg\}\,e^{-\tau}\,.
\end{split}
\end{equation}

\section{Gluon and quark-singlet kernels at LO}
\label{sec:kernels1}

As already advocated, we isolate all singular parts from the kernels
(cf.\ Eq.\ (\ref{pik}))
\begin{equation}
  P_{IK}(t,z)= \frac{1}{(1-z)_+}\delta_{IK} A_{KK}(t)\,+
             \delta(1-z) \delta_{IK}      B_{KK}(t)
	   \, +\,\frac{1}{z}                  C_{IK}(t)
	                                 \,+\,D_{IK}(t,z).
\end{equation}
After expanding to LO (see Eq.\ (\ref{abcd})) the resulting components
$A^{(0)},B^{(0)},C^{(0)},D^{(0)}$ are listed in
Table~\ref{tab:table1}.

\begin{table}[!ht]
\centering
\begin{tabular}{|r|r|r|r|r|r|r|}
\hline
$IK$& $A^{(0)}_{KK}$& $B^{(0)}_{KK}$& $C^{(0)}_{IK}$& $D^{(0)}_{IK}(z)$& $\hat{D}_{IK}(z)$
                                                           & $\int\limits dz  D^{(0)}_{IK}(z)$ \\
\hline
$GG$ & $2C_A$  & $\frac{11}{6}C_A-\frac{2}{3}T_f$  
                           &   $2C_A$ & $2C_A(-2+z-z^2)$   & $0$ &  $-\frac{11}{3}C_A$\\
$qG$ &    $-$   &      $-$ &      $0$ & $2T_f(z^2+(1-z)^2)$& $2T_f$ &$\frac{4}{3}T_f$\\
$qq$ & $2C_F$ & $\frac{3}{2}C_F$   
                           &      $0$ & $C_F(-1-z)$        & $0$ &  $-\frac{3}{2}C_F$\\
$Gq$ &    $-$   &      $-$ &   $2C_F$ & $C_F(-2+z)$        & $0$ &  $-\frac{3}{2}C_F$\\
\hline
\end{tabular}
\caption{\small\sf
  The elements of the singlet LO kernels ($T_f=n_f T_R$).
}
\label{tab:table1}
\end{table}

Let us consider evolution of the simple two-component state
consisting of the gluon and the (singlet) quark
with the LO evolution kernel
\begin{equation}
\Pbf^{(0)}(z)=  
\left[ \begin{matrix}
         P^{(0)}_{GG}(z), & P^{(0)}_{Gq}(z) \\ 
         P^{(0)}_{qG}(z), & P^{(0)}_{qq}(z)
  \end{matrix} \right],
\end{equation}
where
\begin{equation}
\begin{split}
   P^{(0)}_{qG}(z)&=   n_f P^{(0)}_{FG}(z)\,,\\ 
   P^{(0)}_{qq}(z)&=   P^{(0)}_{FF}\,,\\
   P^{(0)}_{Gq}(z)&=   P^{(0)}_{GF}\,,\\
\end{split}
\end{equation}
and $P^{(0)}_{GG}$,   $P^{(0)}_{FG}$, $P^{(0)}_{FF}$,  $P^{(0)}_{GF}$ are given 
explicitly in Appendix~\ref{sec:kernelsNLO}.

Let us concentrate now on the second (less standard) Markovian algorithm
described in Section~\ref{sec:NonStandardMarkov}.
The definition of the simplified kernel matrix elements, Eq.\
(\ref{eq:KernSimp1})  
\begin{equation}
\begin{split}
\hat{\Peu}^\Theta_{IK}(t_0,z)
   &= \Theta(z-\epsilon')\Theta(1-z-\hat\epsilon)
       \frac{\alpha_s(t_0)}{\pi} 
       \bigg\{
             \frac{1}{1-z} \delta_{IK} A^{(0)}_{KK}
           +\frac{1}{z}                  C^{(0)}_{IK}
	                                +\hat{D}_{IK}\bigg\}
\end{split}
\end{equation}
needs $\hat{D}_{IK}$, which are also provided in Table~\ref{tab:table1}.

In this simple case we may explicitly list 
the parton $K\to I$ transition rates
\begin{equation}
\hat\pi_{IK} 
             =\frac{\alpha_s(t_0)}{\pi} \bigg[
	       \delta_{IK} A^{(0)}_{KK} \ln\frac{1}{\hat\epsilon}
	      +            C^{(0)}_{IK} \ln\frac{1}{\epsilon'}
	      +\hat{D}_{IK}
               \bigg]
\end{equation}
as follows
\begin{equation}
  \left[ \begin{matrix} 
        \hat\pi_{GG}, & \hat\pi_{Gq} \\
        \hat\pi_{qG}, & \hat\pi_{qq}
           \end{matrix}\right]
  =\frac{\alpha_s(t_0)}{\pi}
  \left[ \begin{matrix} 
         2C_A\bigg( \ln\frac{1}{\hat\epsilon}+\ln\frac{1}{\epsilon'} \bigg),  
       & 2C_F       \ln\frac{1}{\epsilon'}\\
         2 T_f,
       & 2C_F       \ln\frac{1}{\hat\epsilon}
           \end{matrix}\right]\,,
\end{equation}
and also the characteristic decay rates
$R_K= \sum_{X} \hat\pi_{XK},\; K=g,q $ (in a primary MC)
\begin{equation}
\begin{split}
  R_G&= \frac{\alpha_s(t_0)}{\pi} \bigg\{
         2C_A\bigg( \ln\frac{1}{\hat\epsilon}+\ln\frac{1}{\epsilon'} \bigg)
        +2 T_f
       \bigg\},\\
  R_q&= \frac{\alpha_s(t_0)}{\pi} \bigg\{
         2C_F       \ln\frac{1}{\epsilon'}
        +2C_F       \ln\frac{1}{\hat\epsilon}
       \bigg\}.
\end{split}
\end{equation}

\section{LO kernels for parton-momentum distributions}
\label{sec:kernels2}
\begin{table}[!ht]
\centering
\begin{tabular}{|r|r|r|r|r|r|r|r|}
\hline
$IK$    & $A^{(0)}_{KK}$ & $B^{(0)}_{KK}$ &$F^{(0)}_{IK}(z)$&$\max F^{(0)}_{IK}(z)$
                                                         &$\int\limits F^{(0)}_{IK}(z)dz$\\
\hline
$G\from G$   & $2C_A$    & $\frac{11}{6}C_A-\frac{2}{3}T_f$  
                                            &$2C_A z(-2+z-z^2)$ &   $0$ & $-\frac{11}{6}C_A$\\
$q\from G$      &    $0$          &  $0$          &$T_Rz(z^2+(1-z)^2)$& $T_R$ & $\frac{1}{3}T_R$\\
$\bar{q}\from G$&    $0$          &  $0$          &$T_Rz(z^2+(1-z)^2)$& $T_R$ & $\frac{1}{3}T_R$\\
\hline
$G\from q$      &    $0$          &  $0$          & $ C_F(  2-2z+z^2)$& $2C_F$ & $ \frac{4}{3}C_F$\\
$q\from q$   &  $2C_F$   & $\frac{3}{2}C_F$       & $ C_F( -2 -z-z^2)$&    $0$ & $-\frac{17}{6}C_F$\\
$\bar{q}\from q$&    $0$          &  $0$          &      $0$          &  $0$   &  $0$   \\      
\hline
$G\from\bar{q}$      &    $0$          &  $0$    & $ C_F(  2-2z+z^2)$& $2C_F$ & $ \frac{4}{3}C_F$\\
$q\from\bar{q}$      &    $0$          &  $0$    &      $0$          &  $0$   &  $0$   \\      
$\bar{q}\from\bar{q}$&  $2C_F$ & $\frac{3}{2}C_F$& $ C_F( -2 -z-z^2)$&    $0$ & $-\frac{17}{6}C_F$\\
\hline
\end{tabular}
\caption{\small\sf
  The elements of the LO kernels for parton-momentum distributions ($T_f=n_f T_R$).
}
\label{tab:table2}
\end{table}

We may decompose the evolution kernels for parton-momentum distributions
in the LO and NLO as follows, see Eq.\ (\ref{zpeu})
\begin{equation}
  zP_{IK}(t,z)= \frac{1}{(1-z)_+}\delta_{IK} A_{KK}(t)\,+\,
             \delta(1-z) \delta_{IK} B_{KK}(t)    \,+\,F_{IK}(t,z),
\end{equation}
where $q$ represents one quark flavour and
the corresponding LO components are listed in Table~\ref{tab:table2}.
Comparing to the previous appendix, we have
\begin{equation}
  F_{IK}(t,z) = zD_{IK}(t,z)+C_{IK}(t)-\delta_{IK} A_{KK}(t)\,.
\end{equation}

In the MC we use the simplified
kernels (Eq.\ (\ref{barpcal}))
\begin{equation}
\begin{split}
\bar\Pcal^\Theta_{IK}(\tau_0,z)
   &= \theta({1-z-\bar\epsilon})
       \frac{\alpha_s^{(0)}(t_0)}{\pi} 
       \bigg\{  \frac{1}{1-z}\, \delta_{IK} A_{KK}^{(0)}
	       +F_{IK}^{(0)}(z)\bigg\}.
\end{split}
\end{equation}

In this simple case we may explicitly list 
the parton transition $K\to I$ rates (Eq.\ (\ref{barpi}))
\begin{equation}
\bar\pi_{IK}(\tau_0)
       = \int\limits_0^{1} dz\;
         \bar\Pcal^\Theta_{IK}(\tau_0,z)
       =\frac{\alpha_s^{(0)}(t_0)}{\pi} \bigg[
	 \delta_{IK} A_{KK}^{(0)} \ln\frac{1}{\bar\epsilon}
	 +f_{IK}^{(0)}
         \bigg],
\end{equation}
where 
\begin{equation}
  f_{IK}^{(0)} \equiv \int\limits_0^1 dz\; F_{IK}^{(0)}(z)
\end{equation}
are listed in the Table~\ref{tab:table2}.

The momentum-conservation sum rule reads
\begin{equation}
\sum_X \int\limits_0^{1} dz\; zP_{XK}(t,z)
 =
   B_{KK}(t)
   +\sum_X  \int\limits_0^1 dz\; F_{XK}(t,z) =0\,,
\end{equation}
which is manifest in Table~\ref{tab:table2} for the LO case but it also holds
for the $\overline{MS}$ NLO kernels.
Note that the above sum rule determines unambiguously the virtual
part of the kernels
\begin{equation}
  B_{KK} = -\sum_X  f_{XK} = -\sum_X  \int\limits_0^1 dz\; F_{XK}(z),
\end{equation}
both in the case of the LO and the NLO.

\section{Generic discrete Markovian process}
\label{sec:discrete}

\subsection{Diffusion and evolution equations}

Let us consider a general ``evolution equation'' for the multistate 
{\em discrete} system
\begin{equation}
\label{eq:evolgen}
  \partial_t N_I(t) = \sum_L P_{IL}(t) N_L(t).
\end{equation}
Let us ask whether the time dependence of the above system can {\em always}
be interpreted (implemented) as a probabilistic stochastic Markovian process,
i.e. in terms of MC events with weight equal to $1$.
We shall see that this is not true in the general case and we shall show
under which restriction on the transition matrix $P_{IL}$
the above conjecture on Markovianization is true.
Weighted MC events are excluded from the consideration,
i.e. by the Markovian process we understand 
the probabilistic process with weight=1 events.
Without any loss of generality, in our starting Eq.~(\ref{eq:evolgen})
we have chosen a discrete system in order to simplify the reasoning.

An answer is found by examining a general
``diffusion'' process in the discrete space.
We shall derive the evolution equation (\ref{eq:evolgen}) finding
restriction on the transition matrix $P_{IL}$.
Following this path of reasoning we first define a general transition
probability of an object 
which is exactly in the state $I$ at the initial time $t_0$.
The explicit {\em transition probability} $(t_0,I)\to (t_1,K)$
at the later time $t_1>t_0$ into another state $K\neq I$
is defined as follows
\begin{equation}
  dp(K,t_1|I,t_0)\big|_{K\neq I} 
  = \theta(t_1-t_0)\; P_{KI}(t_1)\; dt_1\;
    e^{ -\int\limits_{t_0}^{t_1} dt' \sum\limits_{J\neq I} P_{JI}(t') }.
\end{equation}
It is properly normalized by the construction, i.e.\ it must fulfill
\begin{equation}
  \int\limits_{t_1>t_0}\; \sum_{K\neq I}\; dp(k,t_1|I,t_0) \equiv 1\,,
\end{equation}
for an {\em arbitrary} starting point $(t_0,I)$.
The probability that the transition to any other state
occurs {\em before} some time $t$ is obtained as
\begin{equation}
  \int\limits_{t_0<t_1<t}\; \sum_{K\neq I}\; dp(K,t_1|I,t_0)
  = 1- e^{ -\Phi_{I}(t,t_0)}\,,
\end{equation}
where we have denoted
\begin{equation}
  \Phi_{I}(t,t_0)= \int\limits_{t_0}^{t} dt' \sum\limits_{J\neq I} P_{JI}(t').
\end{equation}
The probability that such a transition occurs {\em after} the time $t_1=t$
is just equal to
\begin{equation}
  \int\limits_{t_1>t}\; \sum_{K\neq I}\; dp(k,t_1|I,t_0)
  = e^{ -\Phi_{I}(t,t_0)}  \,.
\end{equation}
All the above is a repetition of the very standard description of the
Poissonian ``decay mechanism'' with the ``time-dependent''
transition (decay) constants $P_{JI}(t)>0$; it
describes precisely what we basically do understand as a Markovian process%
\footnote{In the literature one may find many different
  definitions of the Markovian process, 
see e.g.~\cite{Walecka,Kampen,Gardiner}.}.

Let us now imagine a very large ensemble of identical objects,
each of them at a given time $t$ in one well defined state $k$,
and evolving statistically {\em independently} according to the
transition probability distribution defined above.
Let us introduce the population $N_I(t)$ of the objects which are
at a given time $t$ in the state $I$.
Given the above probabilistic transition rule, we may easily calculate
the change of the population $\Delta N_I(t)$ from the time $t$ to the time $t+\Delta t$.
The {\em original} population $N_I(t)$ is diminished to 
$N_I(t) e^{-\Phi_{I}(t+\Delta t,t)}$.
At the same time interval $\Delta t$ the population 
of the state $I$ is also increased by the influx from all other
states $L\neq I$ by $\sum_{L\neq I} P_{IL} N_L \Delta t $.
Altogether we get
\begin{equation}
  N_I(t+\Delta t )= N_I(t) e^{-\Phi_{I}(t+\Delta t,t)}
                   +\sum_{L\neq I} P_{IL}(t) N_L(t) \Delta t\,,
\end{equation}
and 
\begin{equation}
  \Delta N_I(t) = N_I(t+\Delta t )-N_I(t)
  = -N_I(t) \sum_{J\neq I} P_{JI}(t)\Delta t 
   +\sum_{I\neq L} P_{IL}(t)  N_L(t) \Delta t\,,
\end{equation}
or equivalently
\begin{equation}
  \partial_t N_I(t) = -\Big( \sum_{J\neq I} P_{JI}(t) \Big) N_I(t)
                    +\sum_{I\neq L} P_{IL}(t)  N_L(t).
\end{equation}
We therefore conclude that the differential evolution equation
(\ref{eq:evolgen}) can only be compatible with the probabilistic
Markovian process if the following property of the
transition matrix is true
\begin{equation}
    P_{II}(t) \equiv -\sum_{J\neq I} P_{JI}(t).
\end{equation}
This is what we shall always assume to be true in the standard Markovian process.

The QCD singlet evolution kernels do not fulfill
the above condition. 
Hence, perfect Markovianization (with weight=1 events)
is not possible in this context
and the use of the weighted events is mandatory,
at least at the internal level of the parton-shower MC.
MC events can always be turned into weight=1 events with the usual rejection methods,
but at some price. 
The remedy is to use $zP_{IK}(z)$, 
which fulfill the above condition, instead of $P_{IK}(z)$.

\subsection{Iterative solution}
For completness let us write down the iterative solution of the
evolution equation 
\begin{equation}
  \partial_t N_I(t) = -R_I N_I(t)
                    +\sum_{K\neq I} P_{IK}(t)  N_K(t),\quad
  R_I \equiv -P_{II}\,,
\end{equation}
in the discrete space.
The above equation can be easily brought to a homogeneous form
\begin{equation}
\label{eq:iter-homog}
\begin{split}
&e^{-\Phi_I(t,t_0)}
     \partial_t \Big( e^{\Phi_I(t,t_0)} N_I(t) \Big)
    = \sum_{K\neq I} P_{IK}(t) N_K(t),\\
&\Phi_I(t,t_0) \equiv \int\limits_{t_0}^{t} dt_1\; R_I(t_1),
\end{split}
\end{equation}
which then can be turned into an integral equation
\begin{equation}
\begin{split}
N_I(t)&=e^{-\Phi_I(t,t_0)} N_I(t_0)
  +\int\limits_{t_0}^{t} dt_1\;
   e^{-\Phi_I(t,t_1)}
   \sum_{K} P'_{IK}(t_1)\; N_K(t_1)
\\
P'_{KJ} &\equiv \Bigg\{
           \begin{matrix} P_{KJ}, &  \hbox{for~~~} K\neq J, \\
                          0,         &  \hbox{for~~~} K= J, \\
           \end{matrix}
\end{split}
\end{equation}
and finally can be solved by means of multiple iteration
\begin{equation}
  \label{eq:IterBasic2}
  \begin{split}
  N_K(t) = e^{-\Phi_K(t,t_0)} N_K(t_0)
  +&\sum_{n=1}^\infty \;
   \sum_{K_0,\ldots,K_{n-1}}
      \prod_{i=1}^n \bigg[ \int\limits_{t_0}^t dt_i\; \Theta(t_i-t_{i-1}) \bigg]
\\ &\times
      e^{-\Phi_K(t,t_n)}
      \prod_{i=1}^n 
          \bigg[ P'_{K_iK_{i-1}}(t_i) 
                 e^{-\Phi_{K_{i-1}}(t_i,t_{i-1})} \bigg]
      N_{K_0}(t_0),
  \end{split}
\end{equation}
where $K\equiv K_n$, for the brevity of the notation.
The above series of integrals with positively defined integrands
(assuming $P_{IK}\geq 0$)
can be interpreted in terms of a random Markovian
process starting at $t_0$ and continuing until $t$.


\vspace{10mm}
\noindent
{\bf Acknowledgments}\vspace{2mm}\\
We would like to thank P.~Stephens and Z.~Was for the useful discussions.
We acknowledge the warm hospitality of the CERN Physics Department where part
of this work was done. KGB acknowledges the grant from the Polish State
Committee for Scientific Research No. 1 P03B 028 28.


\end{document}